\magnification=\magstep1  
\tolerance500 
\centerline{\bf The Classical Electron
Problem}   
\bigskip 
\centerline{\bf Tepper L. Gill$^{1,2,4}$, W. W. Zachary$^{1,5}$
and J.Lindesay$^{3}$} 
\medskip 
\centerline{{$^{1}$}Department  of  Electrical  Engineering} 
\centerline{{$^{2}$}Department  of  Mathematics}
\centerline{{$^{3}$}Department of Physics}  
\centerline{Howard University,Washington, DC 20059}  
\centerline{E-mail tgill@howard.edu}  
\medskip
\centerline{{$^{4}$}Department of Physics}  
\centerline{University of Michigan}
\centerline{Ann Arbor, Mich. 48109}  
\medskip  
\centerline{{$^{5}$}Department of Mathematics and Statistics}  
\centerline{University of Maryland University College}
\centerline{College Park, Maryland 20742} 
\centerline{E-mail: wwzachary@earthlink.net} 

\medskip 
\noindent  {\bf Abstract}
  
\baselineskip = 1\baselineskip 
\smallskip
\noindent
In this paper, we  construct a parallel image of the conventional Maxwell theory by
replacing the observer-time by the proper-time of the source.  This formulation is
mathematically, but not physically, equivalent to the conventional form.  The change
induces a new symmetry group which is distinct from, but closely related to the
Lorentz group, and fixes the clock of the source for all observers.   The new wave
equation  contains an additional term (dissipative), which arises instantaneously with
acceleration.  This shows that the origin of radiation reaction is not the action of a
``charge" on  itself but arises from inertial resistance to changes in motion. This
dissipative term is equivalent to an effective mass so that classical radiation has both a
massless and a massive part.    Hence, at the local level the theory is one of particles
and fields but there is no self-energy divergence (nor any of the other problems).  
We also show that, for any closed system of particles, there is a global inertial frame
and unique (invariant) global proper-time (for each observer) from which to observe
the system.  This global clock is intrinsically related to the proper clocks of the
individual particles and provides a unique definition of simultaneity for all events
associated with the system.  We suggest that this clock is the historical clock of
Horwitz, Piron, and Fanchi.  At this level, the theory is of the action-at-a-distance type
and the absorption hypothesis of Wheeler and Feynman follows from global
conservation of energy. 

\medskip  
\noindent  PACS classification codes: 03.30.+,03.50.De  \smallskip

\vfill\eject

\baselineskip = 2\baselineskip 

\noindent {\bf 1.0 Introduction}

\smallskip 
\noindent  It was 1865 when James Clark Maxwell published his theory of
electrodynamics.  The slow but steady progress made by our understanding and use of
mechanics and thermodynamics was given a major boost by Maxwell's theory made
practical.  For example, starting from 1866, a continuous communications link has
existed between Europe and the US ( due in no small part to the efforts of Lord
Kelvin).  By 1883, Edison had a workable light bulb, while Bell invented the
telephone in 1886.  The radio waves predicted by Maxwell were discovered by Hertz
in 1887,  and electricity, producing new inventions weekly, was well on the way to
providing what we now consider normal.  

In the intervening 41 years between Maxwell and the introduction of the special
theory of relativity in 1905, a scientific and technological revolution had taken firm
roots.  Indeed, it has been suggested by Feynman$^{1}$ that, `` From the long view of
the history of mankind- seen from, say ten thousand years... there can be little doubt
that the most significant event of the 19th century  will be judged as Maxwell's
discovery of the laws of electrodynamics."  

When the founding fathers, Lorentz, Poincar{\' e}, Einstein, and their contemporaries
began to study the issues associated with the foundations of electrodynamics; they had
a number of options open to them in addressing the fact that the Newtonian theory and
the Maxwell theory were invariant under different transformation groups: (see
Jackson$^{2}$ )
  
\medskip  {\sl
	{\bf 1.} Both theories are incorrect and a correct theory is yet to be found. \smallskip
	{\bf 2.} The ``proper" Maxwell theory will be invariant under the Galilean group.
\smallskip
	{\bf 3.} The ``proper" Newtonian theory will be invariant under the Lorentz group.
\smallskip
	{\bf 4.} The assumption of an ether for electromagnetic propagation is correct  so that 

\quad  Galilean relativity applies to mechanics while electromagnetism has a pre-

\quad ferred reference frame.}
  
\medskip  
At the time, it was unthinkable that the Maxwell theory had any serious
flaws.  Lorentz$^{3,4}$  had recently shown that all of the macroscopic phenomena
of electrodynamics and optics could be accounted for based on an analysis of the
microscopic behavior of electrons and ions.

Einstein$^5$ rejected the fourth possibility and, as noted by Spencer and Shama$^6$,
was the `` first scientist with the foresight to realize that a  formal postulate on the
velocity of light was necessary."  He proposed that all physical theories should satisfy
the (now well-known) postulates of special relativity:
 
\medskip  {\sl
	{\bf 1.} The physical laws of nature and the results of all experiments are independent
 
\quad of the 	particular inertial frame of the observer (in which the experiment is

\quad performed).
   
\medskip
	{\bf 2.} The speed of light in empty space is constant and is independent of the
motion 

\quad of the 	source or receiver.}  
\medskip  
\noindent The first postulate abandons the
notion of an absolute space , while the second abandons absolute time.  In a later
paper, Einstein$^7$ modified the second postulate to make it explicit that he always
referred to observers in inertial frames:  
\medskip  {\bf 2}$'$. {\sl The speed of light in empty space is constant and
independent of the

\quad motion of the 	source or receiver in any inertial frame.}  
\medskip  
Einstein formulated his theory in the usual three-dimensional notation, making a
distinction between time and space.  It was noted by Poincar{\' e}$^8$  that the
transformations of Lorentz could be treated as rotations if time is made an imaginary
coordinate.  Poincar{\' e} had also introduced the metric now attributed to
Minkowski$^9$.   

Although Poincar{\' e} discovered the proper-time, it was Minkowski who recognized
its importance in physical theory and showed that it is the only unique variable
associated with the source and  available to all observers.  Motivated by philosophical
concerns, he further proposed that space and time should not be treated separately, but
should be unified in the now well-known fashion leading to Minkowski space.  Given
the tremendous impact of the then-recent work in geometry on science, it was natural
for him to think along these lines.  Once he accepted this approach, it was also natural
to assume that the proper-time of the source be used to parameterize the motion,
acting as the metric for the underlying geometrization of the special theory of
relativity, thus implicitly requiring that another postulate be added:
 
 \medskip
	{\bf 3.} {\sl The correct implementation of the first two postulates requires that time
be 

\quad treated as a fourth coordinate, and the relationship between components so

\quad constrained to satisfy the natural invariance induced by the Lorentz group of

\quad electrodynamics, (Minkowski space).}  

\medskip  
\noindent  
The four-geometry
postulate was very popular at the time and was embraced by many; but other
important physical thinkers, including Einstein, Lorentz,  Poincar\' e, and Ritz,
regarded it as a mathematical abstraction lacking physical content and maintained that
space and time have distinct physical properties. Although Einstein demurred, the
feeling among many of the leading physicists at that time was that an alternative
implementation should be possible which preserves some remnant of an absolute time
variable (true time), while still allowing for the constancy of the speed of light.  It was
noted by Whittaker$^{10}$  that a few weeks before he died, Lorentz is reported to 
have maintained his belief in the existence of this ``true time".   Dresden$^{11}$
reports that  ``$\cdots$ He retained his beliefs in a Euclidean, Newtonian space time,
and in absolute simultaneity $\cdots$." 
 
\bigskip  
\noindent {\bf 1.1 Perspective} 
\bigskip 
\noindent  The general focus on, and excitement about, the four-geometry left
little room for serious alternative investigations (separated from philosophical
debates).  This is unfortunate since the diversion is part of the reason that the physical
foundations of classical electrodynamics did not receive the early intense investigation
accorded mechanics.  Possibly because of the apparent completeness of the special
theory, interest in statistical mechanics, quantum theory, and the problem of
accelerated motion (the general theory), Einstein was preoccupied with these other
important areas.  On the other hand, the physics community lost three important
thinkers on the subject by 1912.  Ritz died in 1908, Minkowski died (shortly after his
paper appeared) in 1909, and  Poincar\' e died in 1912.  The First World War began in
1914 and within four years decimated a whole generation.  Furthermore, by 1913
interests had already shifted from electrodynamics to the new quantum theory.  The
longer this investigation into classical electrodynamics was delayed, the more
Minkowski's approach became embedded in the culture of physics, permeating the
foundations for all future theories.  By the time problems in attempts to merge the
special theory of relativity with quantum theory forced researchers to take a new look
at the foundations of classical electrodynamics, the Minkowski approach to the
implementation of the special theory was considered almost sacred.

We are now taking our first steps into the twenty first century, one hundred and
forty-five years later.  Electromagnetism is now in the hands of the engineers,
mathematicians, and philosophers, and much of it is not considered mainstream
physics.  For those who learned physics in the sixties and seventies, 
``electrodynamics seems as old as mechanics".  The continued success of quantum
mechanics and the ``apparent" successes of quantum electrodynamics and the standard
model has made the subject pass\' e.  Today, students study the subject as an
introduction to the special theory, preparation for advanced quantum theory, and as a
simple example of a gauge theory.  From this perspective,  there is no real reason to
believe that the first possibility should be rejected out of hand (i.e., that both the
Newtonian and Maxwell theories could in some way be incorrect).  Such a possibility
is even more likely in light of the fact that the problems facing the early workers are
still with us in one form or another.  Furthermore, additional problems have arisen
from both theory and experiment. 

\bigskip

\noindent{\bf 1.2 Problems}  
\medskip

\noindent {\bf Newtonian Mechanics}  
\smallskip  
\noindent 
Once it was accepted that the "proper" Newtonian theory should be invariant under
the Lorentz group, work on this problem was generally ignored until after World War
Two when everyone realized that the quantum theory did not solve the problems left
open by the classical theory.  In particular, it was first noticed that (at  the  classical 
level) Minkowski's  approach  only works (as expected)  in the  one-particle case.  It
was 1948 when Pryce$^{12}$ showed that the canonical center-of-mass is not the
three-vector part  of a four-vector.  This variable is required for any ``natural"
relativistic many-particle theory.  Virtually  all  research since then has focused on
attempts to avoid this problem while maintaining  use of the proper-time of the
observer  as the fourth coordinate  for Minkowski geometry.

In order to provide a simple approach to the problem encountered by Pryce, let us
consider two inertial observers $X$ and $X'$ with the same orientation.  Assume that
the (proper) clocks of $X$ and $X'$ both begin when their origins coincide and $X'$
is moving with uniform velocity ${\bf v}$ as seen by $X$.  Let two particles, each the
source of an electromagnetic field, move with velocities ${\bf w}_i \,  (i=1,2)$, as
seen by $X$, and ${\bf w}_i' \,  (i=1,2) $, as seen by $X'$, so that:  
$$ {\bf x}'_i={\bf x}_i-\gamma ({\bf
v}){\bf v}t+(\gamma ({\bf v})-1)({\bf x}_i\cdot {\bf v}/\left\| {\bf v} \right\|^2){\bf
v},
\eqno(1.1a) 
$$  
$$ 
{\bf x}_i={\bf x}'_i+\gamma ({\bf v}){\bf v}t'+(\gamma ({\bf
v})-1)\left( {{\bf x}'_i\cdot {\bf v}/\left\| {\bf v}\right\|^2} \right){\bf v},
\eqno(1.1b) 
$$   
with $\gamma ({\bf v})=1/\left[ {1-\left( {{\bf v}/c}\right)^2} \right]^{1/2}$,
represent the spacial Lorentz transformations between the corresponding observers. 
Thus, there is clearly no problem in requiring that the positions transform as
expected.  However, when we try to transform the clocks, we see the problem at once
since we must have, for example, 
 $$
 t'=\gamma ({\bf v})\left( {t-{\bf x}_1\cdot {\bf v}/c^2} \right),\ \ \ \ \ t'=\gamma
({\bf v})\left({t-{\bf x}_2\cdot {\bf v}/c^2}\right).
\eqno(1.2a) 
$$   
This is clearly
impossible except under very special conditions on all other observers.  Furthermore,
if we write down the center-of-mass position ${\bf X}$ and require that it transform
as above, we add another (impossible) constraint on the clock of any other observer. 
Pryce's approach is more abstract (and complicated), but leads to the same result.  

In his 1949 paper, Dirac$^{13}$ observed that we must choose a particular realization
of the Poincar\' e  algebra in order to identify the appropriate variables for theory
formulation.  He showed that there are three possible choices of distinct
three-dimensional hypersurfaces that are invariant under subgroups of the Poincar\' e
group and intersect every particle world-line once; the instant form, the point form,
and the front form.  (It was later shown by Leutwyler and Stern$^{14}$  that there  are
five choices.   However, the other two are not especially interesting.)  The instant form
is best known.  It is based on normal time-evolution and uses spacelike hyperplanes in
Minkowski space; the point form is based on mass hyperboloids; while the front form
is based on null hyperplanes.   

Following Dirac's work, Bakamjian and Thomas$^{15}$ showed that one can
construct a quantizable many-particle theory that satisfies the first two postulates of
Einstein.  However, they suggested that when interaction is introduced, their approach
would not permit both a global theory and provide an invariant particle world-line
description (satisfy the third postulate).  This conjecture was generalized and later
proved  by Currie et al$^{16}$ to the effect that the requirements of Hamiltonian
formulation, (canonical) independent-particle variables, and  relativistic covariance
(i.e the canonical positions transform as geometric coordinates), are only compatible
with  noninteracting particles (The No-Interaction Theorem). There are many 
references on  the subject, but the book by Sudarshan and Mukunda$^{17}$ gives a 
comprehensive  review of the problems and  attempts  to  solve them (up to 1974).  All
attempts have ended in failure for one or more reasons which usually include the
inability to quantize.  

The No-Interaction Theorem led many to suspend the requirement that canonical
positions transform as geometric coordinates and to focus on the construction of the 
``correct many-particle  representation for the Poincar\' e algebra".  However, a very
important (but not well-known) theorem was proved by Fong and Sucher$^{18}$ in
1964 for the quantum case, and by Peres$^{19}$ in 1971 for the classical case:

\smallskip  
\noindent{\bf Theorem 1.0} {\sl (Fong-Sucher-Peres)  Suppose that no
restriction is put on the transformation law of the canonical variables of a
many-particle system.  Then given any Hamiltonian $H$, total momentun ${\bf P}$, 
and angular momentum ${\bf J}$ satisfying:
$$ \matrix{ {dH/dt}=0, \hfill & [H, P_m ]=0, \hfill & [H, J_m ]=0, \hfill \cr
\noalign{\vskip3pt} [P_m, P_n ]=0, \hfill & [J_m, P_n ]=\varepsilon_{mns}P_s, \hfill 
& [J_m, J_n ]=\varepsilon_{mns}J_s, \hfill \cr } 
$$    
\noindent it is always possible to
find a boost generator ${\bf L}$ so that the full set of commutation relations of the
Poincar\' e algebra for the inhomogeneous Lorentz group will be satisfied.}

\medskip  
In order to underscore the importance of this theorem, Peres showed
explicitly how to construct a ``clearly" nonrelativistic Hamiltonian and appropriate
boost generator (along with canonical center-of-mass, total momentum, and angular
momentum).  Thus, this theorem implies that a relativistic classical (or quantum)
many-particle theory requires something else besides the commutation relations for
the inhomogeneous Lorentz group.  On the other hand, this is the only requirement
imposed on us by Maxwell's equations! It follows that, contrary to common belief, our
historical (intellectual) state of affairs is not dictated by the Maxwell theory.  We
conclude that the Minkowski postulate imposes an additional condition on the special
theory (not required by Maxwell's equations), but we are still unable to correctly
account for Newtonian mechanics (after almost a hundred years).   Those willing to
dismiss the issue as arcane should be aware that the same problem also exists for the
general theory.   {\sl Thus, the major problem facing us in the twenty first century is to
construct a quantizable classical theory which satisfies the first two postulates of
Einstein in some reasonable form and includes Newtonian Mechanics}. 
 
\medskip  
\noindent {\bf Interpretation}  
\smallskip 
\noindent  
There are interpretation problems with the Minkowski  approach that  are  not
well-known.  First, it should be noted that the conventional use of the words {\it
coordinate time} tends to obscure the  fact  that this is the proper-time  of  the
observer.  This makes physical interpretation complicated  and strange because one is
required to refer back to the proper-time of the source (or the postulated clock of a
co-moving observer) in order  to acquire  a complete interpretation and analysis of
experiments.  Thus, the ``parameter" (used to define the four-geometry) must also be
viewed as a physically real measurable quantity when the theory is  used for
experimental analysis. At  the classical  level  this  asymmetrical  relationship may be
vexing, but it is not contradictory.   However,  at  the quantum  level this same
problem becomes more fundamental.   At this  level,  the  observer proper-time  is  a 
c-number  that transforms  to an operator under the Lorentz group,  while  the
proper-time   of  the  source  is  an  operator  that   remains invariant (see
Wigner$^{20}$). 
     
\medskip  
\noindent  {\bf Radiation Reaction and the Lorentz-Dirac Equation}
\smallskip 
\noindent  
The problems associated with the radiation of accelerated charged particles, and those
of the Lorentz-Dirac equation are old and well-known.  Two books that have
contributed to a clearer understanding of these basic problems are those of
Rohrlich$^{21}$ and Parrott$^{22}$.   Rohrlich provides a comprehensive study of
the classical theory up to 1965, which includes a nice review of the history. (Those
unaware of the continuing effort to solve the classical electron problem should also
see Rohrlich$^{23}$.)   Parrott's book is both clear and insightful.  (His chapter on
the Lorentz-Dirac equation is unbiased, well done, and should be required reading for
any serious student of the subject.)  The classics, Panofsky and Phillips$^{24}$, and
Jackson$^{2}$ are also important sources of insight and history. The elementary (but
correct) account by Feynman$^{1}$ in volume II of his famous lecture series has
done much to educate those with little or no concern with the foundations.  

The radiation of accelerated charged particles is known to occur instantaneously with
acceleration and its nature has been the object of much speculation (see Wheeler and
Feynman$^{25}$).  The great success of Lorentz in using the Maxwell (field) theory,
along with his aether, to show that all of the  macroscopic electrodynamics and optics
could be derived from a microscopic analysis has done much to foster our faith in the
correctness of the theory.  This success carried with it our first introduction to the
divergences of a field theory. He found that the energy density and the field
momentum for each particle diverges unless the particle has a finite radius.  In
addition, the derived (Lorentz) force law did not provide the appropriate dissipation to
account for the observed radiation.  It was also known that the electromagnetic mass
defined by the electrostatic energy divided by $c^2$ and that defined via the
electomagnetic momentum did not agree, giving the well known $4/3$'s problem (see
Schwinger$^{26}$).     

These problems led to the study of various finite-size models for charged particles
and, in turn, forced serious consideration of the action of one part of a charge on itself
(self-energy) and also required the introduction of extra forces to hold the particle
together (Poincar\' e stresses).  
 
The appearance of the classical divergence difficulties in the quantized theory (along
with a few new ones) led many  to hope that the successful construction of a consistent
classical theory would help to solve the corresponding problems in quantum
electrodynamics.  For this reason, many attempts were made to formulate such a
theory.  The most well-known early attempts are due to Born and Infield$^{27}$,
Dirac$^{28}$, Bopp$^{29}$, and Wheeler and Feynman$^{25}$.  (Less well-known
other attempts are due to Rosen$^{30}$, Podolsky and Schwed$^{31}$, and
Feynman$^{32}$.)  Each ran into problems with quantization and are a part of the
history.   However, the point particle reduction theory of Dirac and the
Wheeler-Feynman approach have special importance.  

The use of particles of finite radius causes serious problems with Lorentz invariance,
so a major advance was made when Dirac constructed a point particle reduction theory
for the Lorentz model.  To do this, he used Maxwell's equations to find the retarded
field of the particle, assuming that at large distances the field only contains outgoing
waves, and then calculated the advanced field assuming that at large distances the field
only contains converging waves.  He then defined half the difference between the
retarded and the advanced fields evaluated at the particle position, multiplied by the
charge, as the force of radiation reaction.  This term was added to the Lorentz force to
provide the appropriate dissipation term (the Lorentz-Dirac equation).   This provided
the same dissipation term obtained by Lorentz (in a nonrelativistic calculation), but
was independent  of the particle radius.  Thus, Dirac produced a point particle theory
while all the other problems remained unchanged, and this is essentially what we have
today.  It should also be noted that point particles of finite mass imply infinite
density.  This was a real problem during Newton's time, but does not appear to cause
problems today.  

Wheeler and Feynman took a different ploy.  They showed that we could use point
particles, obtain the same radiation reaction term as above, and eliminate the
self-energy divergence.  Their approach assumes that the field which acts on a given 
particle arises only from other particles (adjunct field).   They used half the sum of the
retarded and the advanced fields, and assumed that there are sufficiently many
particles in the system to completely absorb all radiation given off from any one of
them (absorption hypothesis).   The theory is of the  action-at-a-distance type and also
eliminates the divergences associated with the energy and momentum densities. 
Unfortunately, the theory could not be  quantized, but this work made it clear that the
action-at-a-distance and field theory approaches are much closer than was generally
expected.  (Indeed, Wheeler and Feynman argued that the two theories are
complimentary views.)

The two best-known problems with the Lorentz-Dirac equation are runaway solutions
and preacceleration.  The equation has solutions for a free particle (with no force) that
can self-accelerate off to infinity.  It was conjectured that these solutions were
eliminated by the asymptotic condition proposed by Haag$^{33}$.  However,
Parrott$^{22}$ (pg. 196) notes that the asymptotic condition is necessary to ensure
conservation of  energy-momentum, but may not be sufficient to eliminate all strange
solutions.  Furthermore, the recent paper of Parrott and Endres$^{34}$ makes this
conjecture doubtful.  It has been  recently shown by Low$^{35}$ that this problem
also shows up at the nonrelativistic quantum level.  Things are better for quantum
electrodynamics (they don't appear), but caution is required as the possible existence
of a Landau-like anomalous pole in the photon propagator or the electron-massive
photon forward scattering amplitude could produce the runaway effect. 

The preacceleration problem arises because the equation is nonlocal in time.  This
means that the particle can accelerate prior to the action of a force.  The problem is
generally ignored with the observation that the natural time
 interval for this effect (say for an electron) is of the order of  $6.2\times
10^{-24}\sec$,  so that no classical particle can enter from a free state into interaction
over such a small time interval. 

These problems have existed for sometime now and no solution seems to be in sight. 
It is clear that the first problem is based on the assumption that the dissipation should
be in the Lorentz force and, since this term is of third order in the position variable, the
difficulty follows.  The second problem can be traced back to the use of advanced
fields which are necessary for the theory (Dirac and Wheeler  and Feynman), and to
get the correct dissipation term. 

\medskip  
\noindent  {\bf Mach's Principle and the 2.7 $^{\circ}$K MBR} 

\smallskip
 \noindent  Today, we know that a unique preferred frame of rest exists
throughout the universe and is available to all observers.  This is the 2.7 $^{\circ}$K
microwave background radiation (MBR) which was discovered by Penzias and
Wilson$^{36}$ in 1965 using basic microwave equipment (by today's standards).
   This radiation  is now known to be highly isotropic with anisotropy limits  set at
$0.001\%$.    Futhermore, direct measurements have been made of the velocity of
both our Solar System and Galaxy through  this radiation (370 and 600 $km/sec$
respectively, see Peebles$^{37}$ ).  One can only speculate as to what impact this
information would have had on the thinking of Einstein, Lorentz, Minkowski,
Poincar\'e, Ritz and the many other investigators of the early 1900's who were
concerned with the foundations of electrodynamics and mechanics.  The importance
of this discovery for the foundations of electrodynamics in our view is that this frame
is caused by {\sl radiation from accelerated charged particles}    (independent of the
various cosmological suggestions).

As noted by Peebles, the MBR does not violate the special theory.  However, general
relativity predicts that at each point we can adjust our acceleration locally to find a
freely falling frame where the special theory holds. In this frame, all observers with
constant velocity are equivalent.  Thus, according to the general theory we have an
infinite family of freely falling frames. Within this context, the Penzias and Wilson
findings show that there is a {\sl unique frame in which both the acceleration and
velocity can be set equal to zero at each point in the universe.}  

As suggested by Rohrlich$^{21}$, ``Mach's principle was originally designed to 
ensure that there is no difference between the rotation of the earth with repect to the
fixed stars or the fixed stars with respect to the earth."  It now appears that the fixed
stars  are not needed and the earth really does rotate.  Our concern with this principle is
associated with the fact that an accelerated charged particle experiences a damping 
force simultaneously with the moment of acceleration (relative to any inertial frame). 
Thus, it appears that a charged particle can be used to identify accelerating frames and
raises the question:  what is a charged particle accelerating with respect to?   Put
another way, charged  particles appear to know when they experience a force.  
Furthermore, even if the force is constant, the  effect cannot be transformed away. 
This is a problem for any theory that seeks to unify electromagnetism with gravity.

\bigskip  
\noindent   {\bf 1.3 Purpose} 
\medskip  
\noindent Dirac$^{41}$  was critical
of the use of Minkowski geometry as fundamental. As late as 1963, he  noted  that
``...the picture with four-dimensional symmetry does not give us the whole situation... 
Quantum theory has taught us that we must take a three-dimensional section of what
appears to our consciousness at one time (an observation), and relate it to another 
three-dimensional section  at  another  time."   In reviewing attempts to merge
gravitation with quantum theory, Dirac goes on to question the fundamental nature of
the four-dimensional  requirement in physics and notes that,  in  some cases, physical
descriptions are simplified when one departs from it.  The real question is: What do
we replace it with that  solves the outstanding problems and has some contact with the
physics we know?

A major part of our strong belief in the fundamental nature of the covariant
Minkowski approach to theory construction is based on the
Feynman-Schwinger-Tomonaga formulation of QED and their great computational
success in accounting for the Lamb shift and the anomalous magnetic moment.    The
correct history is at variance with this belief (see Schweber$^{38}$).   It should  first
be noted that, using noncovariant methods, French and Weisskopf$^{39}$, and Kroll
and Lamb$^{40}$ were the first to get the correct results.  The history of the French
and Weisskopf paper can be found in Schweber and is well worth reading.  Both
Schwinger and Feynman  initially got incorrect results using their covariant
formulation and only after the work of  French and Weisskopf was circulated did they
find their mistakes.   Later, Tomonaga got the correct results but used noncovariant 
methods in the middle of the calculation (see Schweber$^{38}$, pg. 270).

\smallskip  
\noindent In attempting to solve the problems of the classical electron,
almost every possible change has been explored except the Minkowski four-geometry
requirement.   Our purpose in this paper is to carefully study the mathematical and
physical implications that arise when we replace the observer proper-time by the
source proper-time in Maxwell's equations.  In order to see how this is possible, we
first recall Minkowski's definition of the proper-time of a source:
$$  d\tau ^2=dt^2-{1 \over {c^2}}d{\bf x}^2=dt^2\left[ {1-\left( {{{\bf w} \over c}}
\right)^2} \right],\ {\bf w}={{d{\bf x}} \over {dt}},
\eqno(1.3a) 
$$ 
$$ d\tau
^2=d{t'}^2-{1 \over {c^2}}d{{\bf x}'}^2=dt^2\left[ {1-\left( {{{{\bf w}'}\over c}}
\right)^2} \right],\ {\bf w}'={{d{\bf x}'} \over{dt'}}.
\eqno(1.3b)  
$$   
Minkowski was
aware that $d{\tau}$ is not an exact one-form and this observation may have affected
his decision to restrict its use to being a parameter for the four-geometry.  However,
there is an important physical reason why it is not an exact (mathematical) one-form. 
Physically, a particle can traverse many different paths (in space) during any given
$\tau$ interval.  This reflects the fact that the distance a particle can travel in a given
time interval depends on the forces acting on it.  This implies that the clock of the
source carries physical information, and there is no a priori physical reason to believe 
that this information is properly encoded when $\tau$ is used as a parameter.  We
rewrite $(1.3)$ as
$$ 
dt^2=d\tau ^2+{1 \over {c^2}}d{\bf x}^2=(d\tau )^2\left[
{1+\left( {{{\bf u} \over c}} \right)^2} \right],\ {\bf u}={{d{\bf x}} \over
{d\tau}},
\eqno(1.4a)  
$$ 
$$ 
d{t'}^2=d\tau ^2+{1 \over {c^2}}d{{\bf x}'}^2=(d\tau )^2\left[ {1+\left( {{{{\bf
u}'} \over c}} \right)^2} \right],\ {\bf u}'={{d{\bf x}'} \over {d\tau}}. 
\eqno(1.4b)  
$$ 
Thus, another possibility appears (which does give an exact one-form).  In case we
have two or more particles, our new time transformations are replaced by (in the
simplest case)  
$$ a'_i\tau _i=\gamma ({\bf v})[a_i\tau _i-{\bf x}_i\cdot{\bf
v}/c^2],
\eqno(1.2b)  
$$  
where $\tau_i$ is the proper-time of the i-th particle and
$a_i$ and $a_i'$ are terms which depend only on $\tau_i$.  
 
\smallskip  
In Section 2 we construct the invariance group which fixes the proper-time of the
source in the single particle case and then explore some of the physical implications
and interpretations of this approach.  At this level we see that the speed of particles
may be faster than the speed of light.  The physical interpretation is that the mass and
the mean lifetime of unstable particles are now both constant, while the velocity
computed using the clock of the source replaces the velocity computed using the
observer clock.  Thus, as will be seen, there is no contradiction with the second
postulate, only a change in conventions.  The second postulate is shown to always
hold for experiments conducted with the source at rest in the frame of the observer, as
is the case for the Michelson-Morley experiment.   

\smallskip  
In Section 3 we show explicitly that Maxwell's equations have an
equivalent representation which fixes the proper-time of the source for all observers. 
We then prove that this formulation is left covariant under the action of the
proper-time group.  Although the fields have the same transformation properties as the
conventional formulation, both the current and charge densities transform differently. 
In particular, we prove that if the charge density is at rest in any inertial frame then it
is invariant (not just covariant) for all observers.  By example, even in the accelerating
case when the proper velocity is $2c$, the relative velocity of our observers must be a
subtantial fraction of $c$ for them to detect any difference in their measured
properties of the charge distribution.

In this Section we also derive the corresponding wave equations and show that they
contain an additional dissipative term,which arises instantaneously with acceleration.   
By a change of variables, we show that the dissipative term is equivalent to an
effective mass for electromagnetic radiation.  We validate this interpretation by
directly calculating the energy radiated by an accelerated charge in the proper-time
formulation. The radiation formulas obtained are close in form but differ from those
computed via the conventional formulation, but agree in the low-velocity limit.  In
particular, the proper-time theory predicts an additional term for the ${\bf E}$-field
which acts along the direction of motion (longitudinal), proving the validity of our
interpretation of the wave equation.   This result means that in the proper-time
formulation, there is no need to to require that the charge act back on itself in order to
account for radiation reaction.  When we couple this result with the the invariance of
the charge density, we are able to prove that the proper-time theory is independent of
the particle size, structure and geometry.  

\smallskip  
In Section 4 we derive the related versions of the optical Doppler effect
and the aberration of wave vectors.  These two phenomena are both well-known and
ubiquitous.  However, the general forms are usually derived using Lorentz
transformations$^{2,42}$.  Here, we derive them from the proper-time theory, using
the new invariance group.  In addition to the usual terms, we obtain new results
because of the nonlocal frequency effects implied by our theory.  These effects play an
important role in our derivation of the group velocity for electromagnetic waves.  Here
we show that the group velocity is $c$ only when measured in the (rest) frame of the
observer, but will not be $c$ for any other observer moving relative to that frame.  The
new value (in the simplest case) will be either $c+v$ or $c-v$, depending on the
direction of the relative motion.  However, as  will be shown in Section 6, this effect is
in the noise for experiments conducted up to  now because of theory interpretation.

\smallskip  
In Section 5 we formulate a global interacting many-particle theory. With
an eye towards the quantum theory, we require that the change from observer
proper-time to source proper-time be canonical.  This leads to the Hamiltonian which
generates $\tau$ translations.  To accomplish this, we use a representation of the
proper-time that is independent of the number of particles.  We derive our
many-particle theory via the commutation relations for the Poincar${\acute e}$
algebra.  As a side benefit, we show that  the global system has a (unique) proper-time
(avaliable for all observers).  This clock provides a unique definition of simultaneity
for all events associated with the system and is (shown to be) intrinsically related to
the proper-times of the particles (in the system).  From these results, it follows that at
the local level, during interaction, the proper-time group is a nonlinear and nonlocal
representation of the Lorentz group.  On the other hand, at the global level, the
proper-time group differs from the Lorentz group by a scale transformation.   It
follows from the work in this Section and in Section 2 that the group representation
space is Euclidean.   

\smallskip  
In Section 6 we explore the ramifications and implications of our 
formulation and discuss some apparent disadvantages.

\bigskip  
\noindent  {\bf 2.0 Proper-Time Transformations}

\smallskip   
\noindent In this section, we derive the transformations that fix the
proper-time of the source for all observers.  If we set $b^2={\bf u}^2+c^2$, then from
$(1.1)$ and $(1.4)$ we have that 
$$ 
{t=(1/c)\int\limits_0^\tau  b(s)ds}\quad {\rm
and}\quad  t'=(1/c)\int\limits_0^\tau {b(s)}'ds.
\eqno{(2.0)} 
$$  
It follows that $t$ and $t'$ are nonlocal as functions of $\tau$ in the
sense that their values depend on the particular physical history (proper-time path) of
the source.  By the mean value property for integrals, we can find a unique $s(\tau)$
for each $\tau$, $0<s(\tau )<\tau$, such that $u_\tau=u(\tau -s(\tau ))$,  and 
$$
t=(1/c)\int\limits_0^\tau  {b(s)ds}=(\bar b_\tau /c)\tau, 
\eqno(2.1a) 
$$  
$$
t'=(1/c)\int\limits_0^\tau  {b'(s)ds}=(\bar b'_\tau/c)\tau. 
\eqno(2.1b) 
$$ 
It is clear that
this property is observer-independent since  
$$ 
t'=\gamma ({\bf v})(t-{{\bf
x}\cdot{\bf v}}/c^2) \Rightarrow (\bar b'_\tau /c) \tau =\gamma ({\bf v})[(\bar
b_{\tau} /c)\tau -({{\bf x}\cdot{\bf v}}/c^2)].
\eqno(2.2) 
$$  
With a fixed clock for all
observers, we can now develop a theory in which only the spatial coordinates are
transformed.  Using $(2.2)$, the required transformations are  
$$ 
{\bf x}'={\bf
x}-\gamma ({\bf v}\ )(\bar b_\tau /c){\bf v}\tau +(\gamma ({\bf v}\ )-1)({{\bf
x}\cdot{\bf v}}/||{\bf v}||^2){\bf v},
\eqno(2.3a) 
$$ 
$$ 
{\bf x}={\bf x}'+\gamma ({\bf
v}\ )(\bar b'_\tau /c){\bf v}\tau \ +(\gamma ({\bf v}\ )-1)({{\bf x}'\cdot {\bf v}}/||{\bf
v}||^2){\bf v}.
\eqno(2.3b) 
$$ 
From a physical point of view, $(2.3)$ tells us
(explicitly) that observers can only share information about the past position of a
given physical system.  The above approach also gives us the only (presently known)
rational solution to the problem of distant simultaneity.  It is clear that all observers
have the option of using their proper clocks with no hope of agreeing on the time
occurrence of any event  associated with the source.  On the other hand, if each
observer agrees to use the proper clock of the source, we see that they will always
agree on the time occurrence of any event associated with the source.

We now see that $a_i=(\bar b_i/c)$ and $a'_i=(\bar b'_i/c)$ in equation $(1.2b)$. The
unit for $b$ and $b'$ is velocity so that physical interpretation is very important.  It
will arise naturally when we represent Maxwell's equations using the proper-time of
the source. For now, we note that they are related by 
$$ 
b'=\gamma ({\bf v})\left[
{b-{{{\bf u}\cdot {\bf v}} \over c}} \right], \ \ \ \ \ \ \ \ b=\gamma ({\bf v})\left[
{b'+{{{\bf u}' \cdot{\bf v}} \over c}} \right]. 
\eqno(2.4) 
$$

For any vector ${\bf d}$, set 
$$ 
{\bf d}^{\dag} ={\bf d}/\gamma ({\bf v})-(1-\gamma
({\bf v}))\left[ {{\bf v}\cdot {\bf d}/(\gamma ({\bf v}){\bf v}^2)} \right]{\bf v}.
\eqno(2.5) 
$$ 
Then the full set of transformations between observers that fix the
proper-time of the source take the (almost) familiar form 
$$ 
{\bf x}'=\gamma ({\bf
v})\left[ {{\bf x}^{\dag} -({\bf v}/c)\bar b_\tau \tau } \right],\ \ \ \ \ \ \ \ {\bf
x}=\gamma ({\bf v})\left[ {{\bf x}'^{\dag}+({\bf v}/c)\bar b'_\tau \tau }
\right],
\eqno(2.6) 
$$ 
$$ 
{\bf u}'=\gamma ({\bf v})\left[ {{\bf u}^{\dag} -({\bf v}/c)b}
\right],\ \ \ \ \ \ \ \ \ \ \ {\bf u}=\gamma ({\bf v})\left[ {{\bf u}'^{\dag} +({\bf v}/c)b'}
\right],
\eqno(2.7) 
$$ 
$$ 
{\bf a}'=\gamma ({\bf v})\left\{ {{\bf a}^{\dag} -{\bf v}\left[
{{\bf u}\cdot {\bf a}/(bc)} \right]} \right\},\ \ \ {\bf a} = \gamma ({\bf v})\left \{ {{\bf
a}'^{\dag} +{\bf v}\left[ {{\bf u}'\cdot {\bf a}'/(b'c)} \right]} \right\}, 
\eqno(2.8)  
$$
where ${\bf a}$ (${\bf a'}$) is the particle proper-(three) acceleration.  The above
transformations (along with (2.4)) form the proper-time group.  In this formulation, we
now have only one clock as an intrinsic part of the theory.

The above transformations are so close to Lorentz transformations that one might
wonder if any new physics is possible.   Not only is there new physics, as will be seen
later, but just as importantly, there are new physical interpretations of old ideas. For
example, relativistic momentum increase is attributed to relativistic mass increase so
that  
$$ 
{\bf p}=m{\bf w},\ \ \ \ m=m_0[1-w^2/c^2]^{-1/2}.
\eqno(2.9a) 
$$  
In the new interpretation,  
$$ 
{\bf p}=m_0{\bf u},\ \ \ \ {\bf u}={\bf w}[1-w^2/c^2]^{-1/2},
\eqno(2.9b) 
$$  
so there is no mass increase, the (proper) velocity increases. Thus, in
particle experiments the particle will have a fixed mass and decay constant,
independent of its velocity. On the other hand, the particle can have (proper) speeds
larger than the speed of light since its velocity is now interpreted to be ${{d{\bf x}}
\mathord{\left/ {\vphantom {{d{\bf x}} {d\tau }}} \right.
\kern-\nulldelimiterspace}{d\tau }}$.  All cases where time dilation is discussed in the
standard approach are replaced by statements about ${\bf u}$ in the new approach.

Note that the relationship between ${\bf u}$ and ${\bf w}$ can be viewed as dual in
the sense that  
$$ 
{\bf u}={\bf w}[1-{{w^2}/{c^2}}]^{-1/2},
\eqno(2.10) 
$$  
$$ 
{\bf
w} = {\bf u}[1+{{u^2}/{c^2}}]^{-1/2}.
\eqno(2.11a) 
$$  
This relationship was first
derived by Schott$^{43}$ in the famous 1915 paper in which he also derived the
well-known Schott term of classical electrodynamics.  Dividing by $c$ in $(2.11a)$,
we get  
$$ 
{{\bf w} \over c}={{\bf u} \over b}.\eqno(2.11b) $$  
It is easy to show that 
$[1+{{u^2} \mathord{\left/{\vphantom {{u^2} {c^2}}} \right.
\kern-\nulldelimiterspace}{c^2}}]^{1/2}=[1-{{w^2} \mathord{\left/{\vphantom
{{w^2} {c^2}}} \right. \kern-\nulldelimiterspace}{c^2}}]^{-1/2}$.  Expanding both
sides and using $(2.11b)$, we have  
$$ 
[1+{{u^2} \mathord{\left/ {\vphantom {{u^2}
{c^2}}}\right. \kern-\nulldelimiterspace} {c^2}}]^{1/2\ }=1+{1 \over 2}{{u^2}\over
{c^2}}-{1 \over 8}{{u^4} \over{c^4}}+\cdots, 
\eqno(2.12)  
$$  
$$  
[1-{{w^2}
\mathord{\left/ {\vphantom {{w^2} {c^2}}}\right. \kern-\nulldelimiterspace}
{c^2}}]^{-1/2\ }=1+{1 \over 2}{{w^2}\over {c^2}}+{3 \over8}
{{w^4}\over{c^4}}+\cdots, 
\eqno(2.13a) 
$$  
$$ 
[1-{{w^2} \mathord{\left/
{\vphantom {{w^2} {c^2}}}\right. \kern-\nulldelimiterspace}
{c^2}}]^{-1/2}=[1-{{u^2}\mathord{\left/ {\vphantom {{u^2} {b^2}}}\right.
\kern-\nulldelimiterspace} {b^2}}]^{-1/2}=1+{1 \over 2}{{u^2}\over {b^2}}+{3
\over 8}{{u^4} \over {b^4}}+\cdots. 
\eqno(2.13b)  
$$  
Thus, all three expressions
agree in the low-velocity region.  It follows that all the results derived from the
standard implementation of special relativity using ${{\bf w} \mathord{\left/
{\vphantom {{\bf w} c}} \right. \kern-\nulldelimiterspace} c}$ can also be
consistently derived using ${{\bf u} \mathord{\left/{\vphantom {{\bf u} b}} \right.
\kern-\nulldelimiterspace}b}$.  This result will be repeatedly exploited in this paper to
provide an alternative interpretation of much of classical electrodynamics.  The real
question that arises is which of these definitions of velocity is appropriate in the
construction of faithful representations of physical reality.   (see Section 6).

\bigskip 
\noindent  {\bf 3.0 Proper-Time Maxwell Equations }

\bigskip 
\noindent  In order to formulate the corresponding Maxwell theory, we need
the following theorem which is derived from  $(2.0)$ and $(2.6)$:

\medskip 
\noindent   {\bf Theorem 3.1} {\sl The transformation properties of the
derivatives when the  observers use the clock of the source are}:  
$$ 
{1 \over
c}{\partial  \over {\partial t}}={1 \over b}{\partial  \over {\partial \tau }},\ {1 \over
c}{\partial \over {\partial t'}}={1 \over {b'}}{\partial  \over {\partial \tau }},
\eqno(3.1)  
$$  
$$ 
{\nabla =\gamma ({\bf v})\left[ {\nabla' -({\bf v}/cb')(\partial  /
\partial \tau )} \right].} \,\,\,\,\ {\nabla' = \gamma ({\bf v})\left[ {\nabla +({\bf
v}/cb)(\partial / \partial \tau )} \right ],}   
\eqno(3.2)  
$$

\noindent {\bf Proof:}  For each case, we prove the first result.  For the first case, we
use the chain rule so that $(1/c)\partial /\partial t = (1/c)(\partial \tau /\partial t)(\partial
/\partial \tau ).$  Using equation (1.3a) and the fact that $[1 - {\bf w}^2 /c^2 ]^{1/2}  =
[1 + {\bf u}^2 /c^2 ]^{ - 1/2} ,$ we have 
$$ 
(1/c)(\partial \tau /\partial t) = (1/c)[1 -
{\bf w}^2/c^2 ]^{1/2}  = (1/c)[1 + {\bf u}^2 /c^2 ]^{ - 1/2}  =(1/b). 
\eqno(3.3a) 
$$
This gives the first part of (3.1).  To prove the first part of (3.2), we use equation (1.1a)
to get that (with an obvious abuse of notation) 
$$ 
{\partial  \over {\partial {\bf x}}} =
{{\partial {\bf x}'} \over {\partial {\bf x}}}{\partial  \over {\partial {\bf x}'}} +
{{\partial t'} \over {\partial {\bf x}}}{{\partial \tau } \over {\partial t'}}{\partial  \over
{\partial \tau}}.
\eqno(3.3b) 
$$ 
Now note that $(\partial {\bf x}'/\partial {\bf x}) =
\gamma ({\bf v}),\,(\partial t'/\partial {\bf x}) =  - \gamma ({\bf v}){\bf v}/c^2 ,$ and
$(\partial \tau /\partial t') = (c/b').$  Putting these terms in equation (3.3b) gives our
result.

We can now formulate the proper-time version of Maxwell's equations.  The
conventional form of these equations for two observers is (in Gaussian units): 
$$
\nabla \cdot {\bf B}=0,\ \ \ \ \ \ \ \ \ \nabla \times {\bf E}+{1 \over c}{{\partial {\bf B}}
\over {\partial t}}=0, 
\eqno(3.4a) 
$$ 
$$ 
\nabla \cdot {\bf E}=4\pi \rho ,\ \ \ \ \nabla
\times {\bf B}={1 \over c}\left[ {{{\partial {\bf E}} \over {\partial t}}+4\pi \rho {\bf
w}} \right], 
\eqno(3.4b) 
$$ 
$$ 
\nabla' \cdot {\bf B}'=0,\ \ \ \ \ \ \ \ \ \nabla' \times {\bf
E}'+{1 \over c}{{\partial {\bf B}'} \over {\partial t'}}=0, 
\eqno(3.5a) 
$$ 
$$ 
\nabla'
\cdot {\bf E}'=4\pi \rho',\ \ \ \nabla' \times {\bf B}'={1 \over c}\left[ {{{\partial {\bf
E}'}\over {\partial t'}}+4\pi \rho' {\bf w}'} \right].
\eqno(3.5b) 
$$ 
Using $(2.11)$ and
$(3.1)- (3.2)$, the above equations can be rewritten using the  proper-time of the
source to get 
$$ \nabla \cdot {\bf B}=0,\ \ \ \ \ \ \ \ \ \nabla \times {\bf E}+{1 \over
b}{{\partial {\bf B}} \over {\partial \tau }}=0,
\eqno(3.6a) 
$$ 
$$ 
\nabla \cdot {\bf
E}=4\pi \rho ,\ \ \ \ \nabla \times {\bf B}={1 \over b}\left[ {{{\partial {\bf E}} \over
{\partial \tau }}+4\pi \rho {\bf u}} \right],
\eqno(3.6b) 
$$ 
$$ 
\nabla' \cdot {\bf B}'=0,\ \
\ \ \ \ \ \ \ \nabla' \times {\bf E}'+{1 \over {b}'}{{\partial {\bf B}'} \over {\partial \tau
}}=0, 
\eqno(3.7a) 
$$ 
$$
 \nabla' \cdot {\bf E}'=4\pi \rho',\ \ \ \nabla' \times {\bf B}'={1
\over {b'}}\left[ {{{\partial {\bf E}'} \over {\partial \tau }}+4\pi \rho' {\bf u}'} \right].
\eqno(3.7b) 
$$

{\it We see that when observers use the proper-time of the source, the velocity of
electromagnetic waves  depends on the motion (of the source), and has magnitude
larger than c. }This may seem strange and even contradictory to the second postulate:
``The speed of light in any inertial frame is constant and is independent of the motion
of the source or receiver." This is not the case.  On closer inspection, it is clear that the
second postulate assumes that the observer's proper-clock is being used to measure
time.  {\it Thus, there is no contradiction, just a change in conventions.}

In the Michelson-Morley experiment, the source is at rest in the frame of the observer
so that ${\bf u}={\bf 0}$ and $b=c$.  It follows that this approach (also) explains the
Michelson-Morley null result.  It also provides agreement with the conceptual (but not
technical) framework proposed by Ritz$^{44}$; namely, that the speed of light does
depend on the (proper) motion of the source.  In this sense, both Einstein and Ritz
were correct.

We could follow Einstein's method$^{5}$ in proving the covariance of the
proper-time equations (using $(3.1)-(3.2)$).  However, we use the four-vector
approach first, to emphasize the fact that our theory is compatible with four-vectors (in
the one-particle case) and second, because it will be convenient for our derivation of
the proper-time transformation of plane waves in Section 4. (The plane waves will be
used to derive formulas for the Doppler effect and aberration of wave vectors.)  
Writing our equations in four-dimensional form as 
$$ 
F=\left[
{\matrix{0&{B_z}&{-B_y}&{-iE_x}\cr {-B_z}&0&{B_x}&{-iE_y}\cr
{B_y}&{-B_x}&0&{-iE_z}\cr {iE_x}&{iE_y}&{iE_z}&0\cr }} \right],\ \ \ {\partial
\over {\partial x_4}}=-{i \over b}{\partial  \over {\partial \tau }},
\eqno(3.8) 
$$ 
it follows that 
$$ 
{{\partial F_{\alpha \beta }} \over {\partial x_\gamma}}+{{\partial
F_{\beta \gamma }} \over {\partial x_\alpha}}+{{\partial F_{\gamma \alpha }}
\over{\partial x_\beta }}=0,\ \ (\alpha ,\beta ,\gamma =1,2,3,4),
\eqno(3.9) 
$$ 
is equivalent to the sourceless equations $(3.4a)$ and 
$$ 
{{\partial F_{\alpha \beta }} \over {\partial x_\beta}}={{4\pi } \over b}J_\alpha,
\quad  J_\alpha =(J_x,J_y,J_z,ib\rho), 
\eqno(3.10) 
$$ 
is equivalent to the proper-time equations with sources $(3.4b)$.  It should be noted
that, in $(3.9)$ and $(3.10)$ and in the sequel, the summation convention is in force
for repeated indices.  If we now define
$[a_{\mu \nu }]$ by 
$$
\left[ {a_{\mu \nu }} \right]=\left[ {\matrix{{1+(\gamma -1)(v_x^2/v^2)}&{(\gamma
-1)[(v_x^{}v_y^{})/v^2]}&{(\gamma
-1)[(v_x^{}v_z^{})/v^2]}&{i\gamma{{v_x^{}} \over c}}\cr {(\gamma -
1)[(v_x^{}v_y^{})/v^2]}&{1+(\gamma-1)(v_y^2/v^2)}& {(\gamma
-1)[(v_y^{}v_z^{})/v^2]}&{i\gamma {{v_y^{}} \over c}}\cr {(\gamma
-1)[(v_x^{}v_z^{})/v^2]}&{(\gamma -1)[(v_y^{}v_z^{})/v^2]}&{1+(\gamma
-1)(v_z^2/v^2)}&{i\gamma{{v_z^{}} \over c}}\cr {-i\gamma {{v_x^{}} \over c}}
&{-i\gamma {{v_y^{}} \over c}}&{-i\gamma {{v_z^{}} \over c}}&\gamma \cr }}
\right], 
\eqno(3.11) 
$$ 
with $\gamma =[1-({v \mathord{\left/ {\vphantom {v
c}}\right. \kern-\nulldelimiterspace} c})^2]^{-1/2}$; then the transformations 
$$
x'_\mu =a_{\mu \nu }x_\nu \ \ (\mu ,\nu=1,2,3,4),  
\eqno(3.12) 
$$  
correspond for $\mu =1,2,3$ to the first set of equations in $(2.6)$ with $x_4=i\bar
b_\tau \tau =i\int_0^\tau  {b(s)ds}$.  Integrating the first equation in $(2.4)$, we have
$$
\int_0^\tau  {b'(s)ds}=\gamma ({\bf v})\left[{\int_0^\tau  {b(s)ds}-{{{\bf x}\cdot {\bf
v}} \over c}}\right]. 
\eqno(3.13) 
$$

Since the transformations $(3.12)$ are equivalent to our proper-time transformations,
we can transform the fields between observers using the four-vector approach just as
is commonly done using Lorentz transformations$^{24,45,46}$.  Thus, we see that
the transformations $F'_{\mu \nu }=a_{\mu \alpha }a_{\nu \beta }F_{\alpha \beta
}^{}\ \ (\mu ,\nu ,\alpha ,\beta =1,2,3,4)$ are equivalent to 
$$ 
{\bf E}'=\gamma \left[
{{\bf E}+{1 \over c}\left({{\bf v}\times {\bf B}} \right)} \right]-(\gamma -1){{({\bf
E}\cdot {\bf v})} \over {{\bf v}^2}}{\bf v},
\eqno(3.14) 
$$ 
$$ 
{\bf B}'=\gamma \left[
{{\bf B}-{1 \over c}\left({{\bf v}\times {\bf E}} \right)} \right]-(\gamma -1){{({\bf
B}\cdot{\bf v})} \over {{\bf v}^2}}{\bf v}. 
\eqno(3.15) 
$$

It should not be surprising that equations $(3.14)$ and $(3.15)$ are the same as would
be obtained if our observers used their own clocks.  This is because the transformation
coefficient matrix $(3.11)$ is the same as is used for Lorentz transformations between
fields.  On the other hand, when we look at the current and charge densities, the
transformations $J_\mu'=a_{\mu \alpha }J_\alpha ^{}\ \ (\mu ,\alpha =1,2,3,4)$ are
equivalent to  
$$ {\bf J}'={\bf J}+(\gamma -1){{({\bf J}\cdot {\bf v})} \over {{\bf
v}^2}}v-\gamma {b \over c}\rho {\bf v},  
\eqno(3.16a)  
$$  
$$ 
b'\rho' =\gamma ({\bf
v})\left[ {b\rho -({{{\bf J}\cdot {\bf v}} \mathord{\left/ {\vphantom {{{\bf J}\cdot
{\bf v}} c}} \right.\kern-\nulldelimiterspace} c})} \right]. 
\eqno(3.16b)  
$$   
Using the first equation of $(2.4)$ in $(3.16b)$, we get:  
$$
 \rho' ={{\rho -({{{\bf
J}\cdot {\bf v}}\mathord{\left/ {\vphantom {{{\bf J}\cdot {\bf v}} {bc}}}\right.
\kern-\nulldelimiterspace} {bc}})}\over {1-({{{\bf u}\cdot {\bf v}} \mathord{\left/
{\vphantom {{{\bf u}\cdot {\bf v}} {bc}}} \right. \kern- \nulldelimiterspace}
{bc}})}}.  
\eqno(3.16c)  
$$  
This result is different from the standard one, (which we obtain if we set $b'=b=c$ in
$(3.16b)$),  
$$ 
\rho' =\gamma ({\bf v})\left[ {\rho
-({{{\bf J}\cdot {\bf v}} \mathord{\left/ {\vphantom {{{\bf J}\cdot {\bf v}} {c^2}}}
\right.\kern-\nulldelimiterspace} {c^2}})} \right]. 
\eqno(3.16d)  
$$  
To see a further difference, if we insert the expression ${{\bf J}/c}={\bf \rho} ({{\bf
u}/b})$ for the current density in $(3.16c)$ and ${\bf J}=\rho {\bf w}$ in $(3.16d)$;
we obtain  
$$
\rho' =\rho {{1-({{{\bf u}\cdot {\bf v}} \mathord{\left/ {\vphantom {{{\bf u}\cdot
{\bf v}} {b^2}}} \right. \kern-\nulldelimiterspace} {b^2}})} \over {1-({{{\bf u}\cdot
{\bf v}} \mathord{\left/ {\vphantom {{{\bf u}\cdot {\bf v}} {bc}}} \right. \kern-
\nulldelimiterspace} {bc}})}}, \eqno(3.17a)  $$  $$ \rho'=\rho \gamma ({\bf v})\left[
{1-({{{\bf w}\cdot {\bf v}} \mathord{\left/ {\vphantom {{{\bf w}\cdot {\bf v}}
{c^2}}} \right.\kern-\nulldelimiterspace} {c^2}})} \right].  
\eqno(3.17b) 
$$ 

In order to obtain a sense of the difference between  $\rho$ and $\rho'$, assume that 
$$
\eqalign{&u=2c\approx u',\ b=\sqrt 5c,\ \Rightarrow w= {\textstyle{2 \over {\sqrt
5}}}c\approx c,\Rightarrow \cr
  &\ \ \ \rho' =\rho \left[ {{{1-{\textstyle{{2v} \over {5c}}}}  \over
{1-{\textstyle{{2v} \over {\sqrt 5c}}}}}} \right],\ \& \ \ \ \ \rho =\rho'
 \left[ {{{1+{\textstyle{{2v} \over {5c}}}} \over {1+{\textstyle{{2v} \over {\sqrt
5c}}}}}} \right].\cr}
$$

It follows that, unless the relative speed of our two observers is a substantial  fraction
of $c$, they will decide that $\rho=\rho'$.  In fact, we obtain the following  remarkable
result from equation $(3.17a)$:

\medskip 
\noindent{\bf Theorem 3.2} {\sl If the source is at rest in the $X$ frame then
$\rho=\rho'$  for all other observers}. 

\smallskip 
\noindent{\bf Proof:} The proof is easy, just note that if ${\bf u=0}$ in
$X$ then  $b=c$ and, from equation $(2.17a)$, $\rho=\rho'$.  Since $X'$  is arbitrary,
the result is true for all observers.

The above theorem means that, in the proper-time formulation, a spherical charge 
distribution at rest in any inertial frame will appear spherical to all other inertial
observers.  As will be shown in the next section, the radiation from an accelerated
charged particle appears as a dissipative term in the wave equations for the fields (i.e.,
neither self-interaction or advanced fields are required).  From these two results, we
see that the proper-time formulation is independent of particle size or structure.

\bigskip 
\noindent   {\bf 3.1 Proper-Time Wave Equations  }

\smallskip 
\noindent If in equations $(3.6)$, we set  
$$ 
{\bf B}=\nabla \times {\bf A},\
\ \ \ \ {\bf E}=-{1 \over b}{{\partial {\bf A}} \over {\partial \tau }}-\nabla \Phi,
\eqno(3.18) 
$$   
then we obtain  
$$ 
\nabla \left[ {\nabla \cdot {\bf A}+{1 \over
b}{{\partial \Phi } \over {\partial \tau }}} \right]+{1 \over b}{\partial \over {\partial
\tau }}\left[ {{1 \over b}{{\partial {\bf A}} \over {\partial \tau }}} \right]- \nabla
^2{\bf A}={1 \over b}\left( {4\pi \rho {\bf u}} \right), 
\eqno(3.19)  
$$  
and 
$$ 
-\nabla
^2\Phi -{1 \over b}{\partial  \over {\partial \tau }}\left[ {\nabla \cdot {\bf A}}
\right]=4\pi \rho.
\eqno(3.20) 
$$  
Imposing the (proper-time) Lorentz gauge  
$$ 
\nabla
\cdot {\bf A}+{1 \over b}{{\partial \Phi } \over{\partial \tau }}=0, 
\eqno(3.21) 
$$  
we get the wave equations 
$$ 
{1 \over {b^2}}{{\partial ^2{\bf A}} \over {\partial
\tau^2}}-{1 \over {b^4}}({\bf u}\cdot {\bf a}){{\partial {\bf A}} \over {\partial \tau
}}-\nabla ^2{\bf A}={1 \over b}\left[ {4\pi \rho {\bf u}} \right], 
\eqno(3.22a) 
$$  
$$
{1 \over {b^2}}{{\partial ^2\Phi } \over {\partial \tau ^2}}-{1 \over {b^4}}({\bf
u}\cdot {\bf a}){{\partial \Phi }\over {\partial \tau }}-\nabla ^2\Phi=4\pi \rho.
\eqno(3.22b) 
$$

We thus obtain a new term that arises because the proper-time of the source carries
information about the interaction that is not available when the proper-time of the
observer is used in formulating theory.   In Section 5 the wave equations will be
derived  for the fields directly to get (no gauge required):  
$$ 
{1 \over
{b^2}}{{\partial ^2{\bf E}} \over {\partial \tau ^2}}-{1 \over {b^4}}({\bf u}\cdot
{\bf a}){{\partial {\bf E}} \over {\partial \tau }}-\nabla^2{\bf E}=-\nabla \left[ {4\pi
\rho {\bf u}} \right]-{1\over b}{\partial  \over {\partial \tau }}\left[ {{{4\pi{\bf J}}
\over b}} \right], 
\eqno(3.23a)  
$$  
$$ 
{1 \over {b^2}}{{\partial ^2{\bf B}} \over
{\partial \tau ^2}}-{1 \over {b^4}}({\bf u}\cdot {\bf a}){{\partial{\bf B}}\over
{\partial \tau }}-\nabla^2{\bf B}={1 \over b}{\partial  \over {\partial \tau}}\left[
{{{4\pi \nabla \times {\bf J}} \over b}} \right]. 
\eqno(3.23b)  
$$  
Thus, the new term is independent of the gauge.  The physical interpretation is clear,
this is a dissipative term which is zero if ${\bf a}$ is zero or orthogonal to ${\bf u}$.
Furthermore, it arises instantaneously with the acceleration of the source.  This is
exactly what one expects of the radiation caused by the inertial resistance of the
source to accelerated motion and is precisely what one means by radiation reaction
(see Wheeler and Feynman$^{25}$).  It should be noted that the creation of real
physical conditions which will make ${\bf a}$ orthogonal to ${\bf u}$ is almost
impossible since ${\bf a}$ arises because of an external force and has no relationship
to ${\bf u}$.  In order to get some insight into the meaning of the new dissipative
terms, let us focus on equation $(3.22b)$.  If we use ${\bf p}=m_0 {\bf u}$ from
equation $(2.9b)$, we see that the external force ${{\bf F}_{ext}}$ satisfies  (This is
only approximate as will be seen in Section 5.4, equation (5.58).)
$$ 
{{\bf F}_{ext}}={{d{\bf p}} \over {d\tau }}=m_0{\bf a}, 
\eqno(3.24a) 
$$  
so that equation $(3.22b)$ becomes 
$$ 
{1 \over {b^2}}{{\partial ^2\Phi } \over
{\partial \tau ^2}}-\left( {{{\bf u} \over b}} \right)\cdot \left( {{{{\bf F}_{ext}} \over
{m_0b^2}}} \right)\left( {{1\over b}{{\partial \Phi } \over {\partial \tau }}} \right)-
\nabla ^2\Phi =4\pi \rho . 
\eqno(3.24b) 
$$ 
If we identify $m_0b^2$ with the effective interaction energy of the particle, then the
middle term can be interpreted as the reactive power loss per unit  interaction energy
of the particle due to its resistance to
${{\bf F}_{ext}}$.  To see this additional term in another physically important way,
use the change of variables $\Phi =\left( {b/c} \right)^{1/2}g$ in $(3.22b)$ to get (see
Courant and Hilbert$^{47}$)  
$$ 
{1 \over {b^2}}{{\partial ^2g} \over {\partial \tau
^2}}- \nabla ^2g+\left[ {{{\ddot b} \over {2b^3}}-{{5\dot b^2} \over {4b^4}}}
\right]g= 4\pi \rho \left({c \over b}\right)^{1/2}. 
\eqno(3.24c) 
$$  
This is the Klein-Gordon equation with an effective mass $\mu$ given by 
$$ 
\mu
=\left\{ {{{\hbar ^2} \over {b^2}}\left[ {{{\ddot b} \over {2b^3}}-{{5\dot b^2} \over
{4b^4}}} \right]} \right\}^{1/2}. 
\eqno(3.25) 
$$ 
Hence, the reactive power loss per unit interaction energy in $(3.24b)$ is equivalent to
an effective mass for the photon that depends on the external force acting on the
particle.

We have only considered our equations at the source. If we look at them in a region
outside the source, there is a major change.  The dissipative term is now constant with
its value fixed at the time the radiation left the source.  Thus, a new picture emerges. 
Every accelerated charged particle emits a continuous stream of (very) small particles
(photons) in all directions.  The energy and the velocity of the particles depend on the
velocity of the source at the moment of emission.  The velocity of the particles
remains constant until they are scattered or absorbed.

\bigskip 
\noindent   {\bf 3.2 Radiation From An Accelerated  Charge }

\bigskip 
\noindent  In this section, we compute the radiation from an accelerated
charge using the proper-time theory.  We can solve  equation $(3.24b)$ directly, but a
better approach is to first find the solution using the proper-time of the observer and
then transform the result to the proper-time of the source. This makes the
computations easier to follow and gives the result quicker.  We follow closely the
approach in Panofsky and Phillips$^{24}$.   In this section, $\left( {{\bf x}(t),t}
\right)$ represents the field position and $\left( {{\bf x}'(t'),t'} \right)$ represents the
retarded position of a point charge source $q$, with  ${\bf r}={\bf x}-{\bf x}'$,
${d{\bf r}}/d{t'}=-{\bf w}$, and ${d^2{\bf r}}/d{t'}^2={\dot{\bf w}}$.  The field
solutions using the standard Lienard-Wiechert potentials are given by 
$$ 
{\bf
A}={{q{\bf w}} \over {cs}},\ \ \ \ \ \Phi ={q \over s},\ \ \ \ s=r-\left( {{{{\bf r}\cdot
{\bf w}} \over c}} \right). 
\eqno(3.26) 
$$ 
The proper-time form is obtained by replacing ${\bf w}/c$ by ${\bf u}/b$ to get 
$$
{\bf A}={{q{\bf u}} \over {bs}},\ \ \ \ \
\Phi ={q \over s},\ \ \ \ s=r-\left( {{{{\bf r}\cdot {\bf u}} \over b}}\right). 
\eqno(3.27)
$$

The field and source-point variables are related by the condition  
$$
 r=\left| {{\bf
x}-{\bf x'}} \right|=c(t-t').
\eqno(3.28) 
$$ 
Here, $d{\bf r}/d\tau '=-{\bf u}=-d{\bf x}'/d\tau '$, where $\tau'$ denotes the retarded
proper-time of the source.  The corresponding {\bf E} and {\bf B} fields can be
computed using equation (3.18) in the form 
$$ 
{\bf E}({\bf x},\tau )={-{1 \over {\bar b}}{{\partial {\bf A}({\bf x},\tau
)}\over {\partial\tau }}-\nabla \Phi({\bf x},\tau ),}\,\,\,{\bf B}({\bf x},\tau )=\nabla
\times {\bf A}({\bf x},\tau )
\eqno(3.29) 
$$ 
with ${\bar {\bf u}}=d{\bf x}/d{\tau}$, where $\tau$ denotes the proper-time of the
present position of the source and $\bar b=\left( {\bar {\bf u}^2+c^2} \right)^{1/2}$. 
In order to compute the fields from the potentials, we note that the components of the
$\nabla$ operator are partials at constant time $\tau$, and therefore are not at constant
$\tau'$.  Also, the partial derivatives with respect to $\tau$ imply constant ${\bf x}$
and hence refer to the comparison of potentials at a given point over an interval in
which the coordinates of the source will have changed.  Since only time variations
with respect to $\tau'$ are given, we must transform $(\partial /\partial \tau )\left|
{_{\bf x}}\right.$ and $\nabla
\left| {_\tau } \right.$ to expressions in terms of ${\partial/{\partial \tau' }}\left| {_{\bf
x}} \right.$.  To do this, we must first transform $(3.28)$ into a relationship between
$\tau$ and $\tau'$.   The  required correspondence is  
$$ 
c(t-t')=\int_{\tau' }^\tau {b(s)ds}.
\eqno(3.30) 
$$

It is easier to first relate ${\partial / {\partial t}}\left|{_{\bf x}}\right.$   to   ${\partial /
{\partial t'}} \left|{_{\bf x}}\right.$ and then convert them to relationships between
${\partial/{\partial \tau }}{\left|{_{\bf x}}\right.}$ and 
${\partial/{\partial\tau'}}{\left|{_{\bf x}}\right.}.$   The following are in reference
24, pg. 298:  
$$ 
{{\partial r} \over {\partial t'}}=-{{{\bf r}\cdot {\bf w}} \over r},\ \
{{\partial r} \over {\partial t}}=c\left( {1-{{\partial t'} \over {\partial t}}}
\right)={{\partial r} \over {\partial t'}}\cdot {{\partial t'} \over {\partial t}}=-{{{\bf
r}\cdot {\bf w}} \over r}{{\partial t'} \over {\partial t}}.
\eqno(3.31) 
$$   
Since ${{\partial \tau }/{\partial t}}={c /b}$, we have  
$$ 
{{\partial r} \over {\partial
t}}=c{\partial  \over {\partial t}}\left( {t-t' } \right)={{\partial \tau } \over {\partial
t}}{\partial  \over {\partial \tau }}\int_{\tau' }^\tau  {b(s)ds}={c \over {\bar b}}\left[
{\bar b-b{{\partial \tau' } \over {\partial \tau }}} \right].
\eqno(3.32)  
$$   
We also have, using ${{\partial \tau' }/{\partial t'}}={c/b}$ , that  
$$ 
{{\partial r} \over
{\partial t'}}={{\partial r} \over {\partial \tau' } }{{\partial \tau' } \over {\partial
t'}}={c \over b}{{\partial r} \over {\partial \tau' }}\Rightarrow {1 \over b}{{\partial
r} \over {\partial \tau' }}=-{{{\bf r}\cdot {\bf w}} \over {rc}}=-{{{\bf r}\cdot {\bf
u}} \over {rb}},
\eqno(3.33)  
$$  
so ${{\partial r} /{\partial\tau' }}=-{{{\bf r}\cdot{\bf u}}/ r}$ and hence 
$$ 
{{\partial r} \over {\partial t}}={{\partial r} \over {\partial \tau
}}{c \over {\bar b}}={c \over {\bar b} }\left[ {\bar b-b{{\partial \tau' } \over {\partial
\tau }}} \right]\Rightarrow {{\partial r} \over {\partial \tau }}=\left[ {\bar
b-b{{\partial \tau' } \over {\partial \tau }}} \right],
\eqno(3.34) 
$$ 
$$ 
{{\partial r} \over
{\partial \tau }}={{\partial r} \over {\partial \tau' }}{{\partial \tau' } \over {\partial
\tau }}=-{{{\bf r}\cdot {\bf u}} \over r}{{\partial \tau' } \over {\partial \tau
}}\Rightarrow \ \ \ -{{{\bf r}\cdot {\bf u}} \over r}{{\partial \tau' } \over {\partial \tau
}}=\left[ {\bar b-b{{\partial \tau' } \over {\partial \tau }}} \right].
\eqno(3.35) 
$$ 
Solving (3.35) for ${{\partial \tau' }/ {\partial \tau }}$, we get 
$$ 
{{\partial \tau' }
\over {\partial \tau }}={{\bar b} \over b}{r \over s},\ \ \ s=r-{{{\bf r}\cdot {\bf u}}
\over b}.
\eqno(3.36) 
$$  
Using this, we see that 
$$ 
{1 \over {\bar b}}{\partial  \over
{\partial \tau }}={1 \over b}\cdot {r \over s}{\partial  \over {\partial \tau'}}.
\eqno(3.37) 
$$   
From $\nabla r=-c\nabla t'=\nabla_1r+({{\partial r}/ {\partial
t'}}){\nabla t'}$, we see that 
$$ 
\nabla r={{\bf r} \over r}-{c \over b}\cdot {{{\bf
r}\cdot {\bf u}} \over r}\nabla t'\ \ \Rightarrow -c\nabla t'={{\bf r} \over r}-{c \over
b}\cdot {{{\bf r}\cdot {\bf u}} \over r}\nabla t'.
\eqno(3.38) 
$$   
Using $c\nabla
t'=b\nabla \tau' $ and solving for $\nabla \tau' $, we get $\nabla \tau' =-\left( {{{\bf r}
\mathord{\left/ {\vphantom {{\bf r} {bs}}} \right. \kern-\nulldelimiterspace} {bs}}}
\right)$, so that 
$$
 \nabla =\nabla _1-{{\bf r} \over {bs}}\cdot {\partial \over {\partial
\tau' }}. 
\eqno(3.39) 
$$    
We now compute ${\nabla _1s}$ and  ${\partial s}/{\partial
\tau' }$.  The calculations are easy, so we simply state the results: 
$$ 
\nabla _1s={{\bf
r} \over r}-{{\bf u} \over b}={1 \over r}\left( {{\bf r}-{{r{\bf u}} \over b}}  \right),
\eqno(3.40) 
$$ 
$$ 
{{\partial s} \over {\partial \tau' }}={{{\bf u}^2} \over b}-{{{\bf
r}\cdot {\bf u}} \over r}-{{{\bf r}\cdot {\bf a} } \over b}+{{\left( {{\bf r}\cdot {\bf
u}} \right)\left( {{\bf u}\cdot {\bf a}} \right)} \over {b^3}}.
\eqno(3.41) 
$$

We can now calculate the fields.  The computations are long but follow those of
reference 24, so we only record a few selected results.  We obtain  
$$
\eqalign{&-\nabla \Phi ={q \over {s^2}}\nabla s={q \over {s^2}}\left( {\nabla
_1s-{{\bf r} \over {bs}}\cdot{{\partial s} \over {\partial \tau }} } \right)\Rightarrow
\cr &-\nabla \Phi ={{q\left[ {{\bf r}\left( {1-{{{\bf u}^2} \mathord{\left/ {\vphantom
{{{\bf u}^2} {b^2}}} \right. \kern-\nulldelimiterspace} {b^2}}} \right)-{{{\bf u}s}
\mathord{\left/ {\vphantom {{{\bf u}s} b}} \right. \kern-\nulldelimiterspace} b}}
\right]} \over {s^3}}+{{q{\bf r}\left( {{\bf r}\cdot {\bf a}} \right)} \over
{b^2s^3}}-{{q{\bf r}\left( {{\bf r}\cdot {\bf u}} \right)\left( {{\bf u}\cdot {\bf a}}
\right)} \over {b^4s^3}}.\cr}
\eqno(3.42) 
$$  
Now  use equation $(3.37)$ to get  
$$
 -{1 \over {\bar b}}{{\partial {\bf A}} \over {\partial \tau }}=\left( {-{1 \over b}}
\right)\left( {{r \over s} } \right){{\partial {\bf A}} \over {\partial \tau' }}\Rightarrow
$$  
$$ 
\eqalign{&-{1 \over {\bar b}}{{\partial A} \over {\partial \tau }}={{-\left(
{qr{\bf u}/b} \right)\left\{ {\left( {{\bf u}/b} \right)\cdot \left[ {\left( {{\bf r}/r}
\right)-\left( {{\bf u}/b} \right)} \right]} \right\}} \over {s^3}}\cr &+{{-qr^2{\bf
a}+qr\left\{ {{\bf r}\times \left[ {{\bf a}\times \left( {{\bf u}/b} \right)} \right]}
\right\}} \over {b^2s^3}}+{{q{\bf u}\left[ {({\bf r}\cdot {\bf r})\left( {{\bf u}\cdot
{\bf a}} \right)} \right]}\over {b^4s^3}}. \cr}
\eqno(3.43) 
$$  
Combining $(3.42)$ and $(3.43)$, we get  
$$ 
\eqalign{&{\bf E}({\bf x},\tau )=-{1 \over {\bar b}}{{\partial
{\bf A}({\bf x},\tau )} \over {\partial \tau }}-\nabla \Phi (x,\tau )\Rightarrow \cr 
&{\bf E}({\bf x},\tau )={{q\left[ {{\bf r}\left( {1-{\bf u}^2/b^2} \right)-{\bf u}s/b}
\right]} \over {s^3}}-{{\left( {-qr{\bf u}/b} \right)\left[ {({\bf u}/b)\cdot ({\bf
r}/r-{\bf u}/b)} \right]} \over {s^3}}\cr  &+{{-q\left[ {r^2{\bf a}-{\bf r}\left( {{\bf
r}\cdot {\bf a}} \right)} \right]+qr\left[ {{\bf r}\times \left( {{\bf a}\times {\bf u}/b}
\right)} \right]} \over {b^2s^3}}+{{q\left( {{\bf u}\cdot {\bf a}} \right)\left[ {{\bf
u}r^2-{\bf r}\left( {{\bf r}\cdot {\bf u}} \right)} \right]} \over {b^4s^3}
}.\cr}
\eqno(3.44)  
$$   
Finally, using standard vector identities and combining terms,
we get (with ${\bf r}_{\bf u}={\bf r}-{\bf u}r/b $)  
$$ 
\eqalign{&{\bf E}({\bf x},\tau
)={{q\left[ {{\bf r}_{\bf u}\left( {1-{\bf u}^2/b^2} \right)} \right]} \over
{s^3}}+{{q\left\{ {{\bf r}\times \left[ {{\bf r}_{\bf u}\times {\bf a}} \right]}
\right\}} \over {b^2s^3}}\cr  &+{{q\left( {{\bf u}\cdot {\bf a}} \right)\left[ {{\bf
r}\times  \left( {{\bf u}\times {\bf r}} \right)} \right]} \over {b^4s^3}}.\cr}
\eqno(3.45) 
$$
The computation of ${\bf B}$ is similar:  
$$ 
\eqalign{&{\bf B}({\bf x},\tau)={{q\left[ {({\bf r}\times {\bf r}_{\bf u})(1-{\bf
u}^2/b^2)} \right]} \over {rs^3}}+{{q{\bf r}\times \{ {\bf r}\times [{\bf r}_{\bf
u}\times {\bf a}]\} } \over {rb^2s^3}}\cr  &+{{qr({\bf u}\cdot {\bf a})({\bf r}\times
{\bf u})} \over {b^4s^3}}.\cr} 
\eqno(3.46)  
$$

It is easy to see that we have ${\bf B}=\left( {{{\bf r} \mathord{\left/ {\vphantom
{{\bf r} r}} \right. \kern-\nulldelimiterspace} r}} \right)\times {\bf E}$ so that ${\bf
B}$ is orthogonal to ${\bf E}$.  The first two terms in $(3.45)$ and $(3.46)$ are the
same as (19-13) and (19-14) in reference 24 (pg. 299).  The last term in each case
arises because of the dissipative terms in equations $(3.22)$ and $(3.23)$.

The last terms in $(3.45)$ and $(3.46)$ are zero if $\bf a$ is zero or orthogonal to $\bf
u$.  In the first case, there is no radiation and the particle moves with constant velocity
so that the field is massless. As noted earlier, the second case depends on conditions
that are impossible in practice, namely the creation of motion which keeps $\bf a$
orthogonal to $\bf u$.   Since  ${\bf r}\times \left( {{\bf u}\times {\bf r}}
\right)=r^2{\bf u}-\left( {{\bf u}\cdot {\bf r}} \right){\bf r}$, we see that there is a
component along the direction of propagation (longitudinal).  Hence, in all other
cases, there is a small mass associated with electromagnetic radiation which varies
with the acceleration of the particle.

\bigskip 
\noindent   {\bf 3.3 Radiated Energy}

\smallskip  
\noindent In light of the difference in the calculated fields, it becomes
important to also compute the radiated energy for the proper-time theory and compare
it with the Minkowski formulation.  It is well-known that the radiated energy is
determined by the Poynting vector, which is defined by ${\bf P}=\left( {{c
\mathord{\left/ {\vphantom {c {4\pi }}} \right. \kern- \nulldelimiterspace} {4\pi }}}
\right)\left( {{\bf E}\times {\bf B}} \right)$.

To calculate the angular distribution of the radiated energy, we must be careful to note
that the rate of radiation is the amount of energy lost by the charge in a time interval
$d\tau'$ during the emission of the signal $\left( {{{-dU} \mathord{\left/ {\vphantom
{{-dU} {d\tau' }}} \right. \kern-\nulldelimiterspace} {d\tau' }}} \right)$.  However (at
a field point), the Poynting vector ${\bf P}$ represents the energy flow per unit time
measured at the present time ($\tau$).   With this understanding, the same approach
that leads to the above formula gives ${\bf P}=\left( {{{\bar b} \mathord{\left/
{\vphantom {{\bar b} {4\pi }}} \right. \kern-\nulldelimiterspace} {4\pi }}}
\right)\left( {{\bf E}\times {\bf B}} \right)$ in the proper-time formulation.  We thus
obtain the rate of energy loss of a charged particle into a given infinitesimal solid
angle $d\Omega $ as 
$$ 
-{{dU} \over {d\tau' }}(\Omega )d\Omega =\left( {{{\bar b}
\mathord{\left/ {\vphantom {{\bar b} {4\pi }}} \right. \kern-\nulldelimiterspace}
{4\pi }}} \right)\left[ {{\bf n}\cdot \left( {{\bf E}\times {\bf B}} \right)} \right]{\bf
r}^2{{d\tau } \over {d\tau'  }}d\Omega .
\eqno(3.47) 
$$  
Using equation $(3.36)$, we get that $\left( {{{d\tau } \mathord{\left/ {\vphantom
{{d\tau } {d\tau' }}} \right. \kern-\nulldelimiterspace} {d\tau' }}} \right)={{bs}
\mathord{\left/ {\vphantom {{bs} {\bar br}}} \right. \kern-\nulldelimiterspace} {\bar br}}$, so that $(3.47)$
becomes 
$$ 
-{{dU} \over {d\tau' }}(\Omega )d\Omega =\left( {{b \mathord{\left/
{\vphantom {b {4\pi }}} \right. \kern- \nulldelimiterspace} {4\pi }}} \right)\left[
{{\bf n}\cdot \left( {{\bf E}\times {\bf B}} \right)} \right]rsd\Omega .
\eqno(3.48) 
$$ 
As is well-known, only those terms that fall off as $\left( {{1 \mathord{\left/
{\vphantom {1 r}} \right. \kern- \nulldelimiterspace} r}} \right)$ (the radiation terms)
in $(3.45)$ and $(3.46)$ contribute to the integral of $(3.48)$.  It is easy to see that our
theory gives the following radiation terms: 
$$
 {\bf E}_{rad}={{q\left\{ {{\bf r}\times
\left[ {{\bf r}_{\bf u}\times {\bf a}} \right]} \right\}} \over {b^2s^3}}+{{q\left( {{\bf
u}\cdot {\bf a}} \right)\left[ {{\bf r}\times \left( {{\bf u}\times {\bf r}} \right)}
\right]} \over {b^4s^3}}={\bf E}_{rad}^c+ {\bf E}_{rad}^d,
\eqno(3.49) 
$$ 
$$ 
{\bf
B}_{rad}={{q{\bf r}\times \left\{ {{\bf r}\times \left[ {{\bf r}_{\bf u}\times {\bf a}}
\right]} \right\}} \over {rb^2s^3}}+{{qr\left( {{\bf u}\cdot {\bf a}} \right)\left( {{\bf
r}\times {\bf u}} \right)} \over {b^4s^3}}={\bf B}_{rad}^c+{\bf
B}_{rad}^d,
\eqno(3.50) 
$$ 
where ${\bf E}_{rad}^c,{\bf B}_{rad}^c$ are of the
same form as the classical terms with $c$ replaced by $b$, ${\bf w}'$ by ${\bf u}$,
and $\dot {\bf w}'$ by ${\bf a}$.  The two terms ${\bf E}_{rad}^d,{\bf B}_{rad}^d$,
are new and come directly from the dissipation term in the wave equations. (Note the
characteristic ${{\left( {{\bf u}\cdot {\bf a}} \right)}\mathord{\left/ {\vphantom
{{\left( {{\bf u}\cdot {\bf a}} \right)} {b^4}}} \right. \kern-\nulldelimiterspace}
{b^4}}$.)  We can easily integrate the classical terms to see that 
$$
\eqalign{&\int\!\!\!\int_\Omega  {\left( {{{-dU^c} \mathord{\left/ {\vphantom
{{-dU^c} {d\tau }}} \right. \kern- \nulldelimiterspace} {d\tau }}} \right)}d\Omega
\cr  &=\left( {{b \mathord{\left/ {\vphantom {b {4\pi }}} \right.
\kern-\nulldelimiterspace} {4\pi }}} \right)\int\!\!\!\int_\Omega  {\left[ {{\bf n}\cdot
\left( {{\bf E}_{rad}^c\times {\bf B}_{rad}^c} \right)} \right]rsd\Omega }={2 \over
3}{{q^2\left| {\bf a} \right|^2} \over {b^3}}.\cr}
\eqno(3.51) 
$$  
This agrees with the standard result for small proper- velocity and proper-acceleration
of the charge when $b\approx c$ and ${\bf a}\approx {{d{\bf w}} \mathord{\left/
{\vphantom {{d{\bf w}} {dt}}} \right. \kern- \nulldelimiterspace} {dt}}$.

In the general case, our theory gives additional effects because of the dissipative
terms.  To compute the integral of $(3.48)$, we use spherical coordinates with the
proper-velocity ${\bf u}$ directed along the positive z-axis.  Without loss of
generality, we orient the coordinate system so that the proper-acceleration ${\bf a}$
lies in the xz-plane.  Let ${\alpha}$ denote the acute angle between ${\bf a}$ and
${\bf u}$, and substitute $(3.49)$ and $(3.50)$ in $(3.48)$ to obtain 
$$
\eqalign{&-{{dU} \over {d\tau }}(\Omega )d\Omega =\cr  &={{q^2\left| {\bf a}
\right|^2} \over {4\pi b^3}}\left\{{} \right.\left( {1-\beta \cos \theta } \right)^{-4}\left[
{1-\sin ^2\theta \sin ^2\alpha \cos \phi } \right.\cr  &\left. {-\cos ^2\theta \cos ^2\alpha
-\left( {{1 \mathord{\left/ {\vphantom {1 2}} \right. \kern- \nulldelimiterspace} 2}}
\right)\sin 2\theta \sin 2\alpha \cos \phi } \right]\cr  &-2\beta \left( {1-\beta \cos \theta
} \right)^{-5}\left( {\sin ^2\theta \cos \alpha -\left( {{1 \mathord{\left/ {\vphantom {1
2}}\right. \kern- \nulldelimiterspace} 2}} \right)\sin 2\theta \sin \alpha \cos \phi }
\right)\chi \cr
  &\left. {+\beta ^2\sin ^2\theta \left( {1-\beta \cos \theta } \right)^{-6}\chi ^2}
\right\},\cr} 
\eqno{(3.52)} 
$$ 
where 
$$
 \chi ={{b^2} \over {r\left| {\bf a}
\right|}}\left( {1- \beta ^2} \right)+\beta \cos \alpha \left( {1-{1 \over \beta}\cos \theta
} \right)-\sin \theta \sin \alpha \cos \phi ,
\eqno(3.53) 
$$   
and $\beta =\left({{{\left| {\bf u} \right|} \mathord{\left/ {\vphantom {{\left| {\bf u}
\right|} b}} \right. \kern-\nulldelimiterspace} b}} \right)$.

The integration of $(3.52)$ over the surface of the sphere is elementary, and we
obtain, after some extensive but easy computations (which are summarized in the
appendix): 
$$ 
\eqalign{&\mathop {\lim }\limits_{r\to \infty }\int\!\!\!\int {-{{dU}
\over {d\tau }}(\Omega )d\Omega }\cr &={2 \over 3}{{q^2\left| {\bf a} \right|^2}
\over {b^3}}\left( {1-\beta ^2} \right)^{-3}\left[ {1-{1 \over 5}\beta ^2\left( {4+\beta
^2} \right)} \right.+\left. {{1 \over 5}\beta ^2\left( {6+\beta ^2} \right)\sin ^2\alpha }
\right].\cr}
\eqno(3.54) 
$$   
As can be seen, this result agrees with $(3.51)$ at the lowest order.  For comparison,
the same calculation using the observer's clock for the case of general orientation of
velocity ${{d{\bf x}'}\mathord{\left/ {\vphantom {{d{\bf x}'} {dt'}}} \right.
\kern-\nulldelimiterspace} {dt'}}$ and acceleration ${{d{\bf w}'}\mathord{\left/
{\vphantom {{d{\bf w}'} {dt'}}} \right. \kern-\nulldelimiterspace} {dt'}}$ is 
$$
\eqalign{&\mathop {\lim }\limits_{r\to
\infty}\int\!\!\!\int {-{{dU} \over {dt}}(\Omega )d\Omega }={2 \over
3}{{q^2\left|{\dot {\bf w}'} \right|^2} \over {c^3}}\left( {1-\beta ^2} \right)^{-
3}\left[ 1 \right.-\left. {\beta ^2\sin ^2\alpha} \right]\cr}, 
\eqno(3.55) 
$$   
where $\beta =\left({{{\left| {\bf w}' \right|} \mathord{\left/ {\vphantom {{\left| {\bf
w}' \right|} c}} \right.\kern-\nulldelimiterspace} c}} \right)$.

We observe that, in general, for an arbitrary angle ${\alpha}$ with ${0\le{\alpha}
\le{{\pi}/2}}$ and arbitrary ${\beta}$ between $0$ and $1$, our result does not agree
with (3.55) even if we replace $b$ with $c$ and ${\bf a}$ with $d{\bf w}'/dt'$. This
shows, along with our other results, that the apparently small change in clocks induces
large changes in the physical predictions.  We will return to this point in the
conclusion of the paper.

\bigskip  
\noindent {\bf 4.0  Proper-Time Doppler Effect and Aberration}

\smallskip 
\noindent  In this section, we apply our proper-time theory to compute the
optical Doppler effect and aberration.  To do this, we first consider the transformation
properties of plane wave solutions to Maxwell's equations.  Assuming that our
observers are in the far-field of the source so that, to a good approximation, the waves
are plane when they arrive at the observers' positions, we want solutions of the form
(${\bf E}_0=const,\ {\bf B}_0=const$) 
$$ 
{\bf E}=\Re \left\{ {{\bf E}_0\exp \left[
{i\left( {{\bf k}\cdot {\bf x}-{1 \over c}\int_0^\tau  {\omega (s)b(s)ds}} \right)}
\right]} \right\},
\eqno(4.1a) 
$$ 
$$ 
{\bf B}=\Re \left\{ {{\bf B}_0\exp \left[ {i\left(
{{\bf k}\cdot {\bf x}-{1 \over c}\int_0^\tau  {\omega (s)b(s)ds}} \right)} \right]}
\right\},
\eqno(4.1b) 
$$  
where, in accordance with equations $(2.0)$, we have modified the plane wave
representations to allow for proper-time (nonlocal) dependence of the frequency. 
Assuming that the frequency is a differentiable function of time, we get that the above
plane wave representations of the fields are solutions of the wave equations in the
far-field region (where the charge and current densities are zero), 
$$ 
{1 \over
{b^2}}{{\partial ^2{\bf E}} \over {\partial \tau ^2}}-{1 \over {b^4}}({\bf u}\cdot
{\bf a}){{\partial {\bf E}} \over {\partial \tau }}-\nabla^2{\bf E}=0, 
\eqno(3.23a) 
$$
$$ 
{1 \over {b^2}}{{\partial ^2{\bf B}} \over {\partial
\tau^2}}-{1 \over {b^4}}({\bf u}\cdot {\bf a}){{\partial {\bf B}} \over {\partial \tau
}}-\nabla^2{\bf B}=0, \eqno(3.23b) $$ provided that $$ {\bf k}^2={{\omega (\tau
)^2} \over {c^2}}\left[ {1+i{{c\dot \omega (\tau )} \over {b\omega (\tau )^2}}}
\right]. 
\eqno(4.2a) 
$$ 
In addition, from $(3.4)$ we have 
$$ 
{\bf k}\cdot {\bf B}_0=0,\ \ {\bf k}\cdot {\bf E}_0=0, 
\eqno(4.2b) 
$$ 
$$ 
{\bf k}\times {\bf E}_0={{\omega (\tau )} \over c}{\bf B}_0. 
\eqno(4.2c) 
$$ 
It follows from $(4.2a)$ that the wave vector
${\bf k}$ depends on $\omega(\tau )$ and its derivative $\dot \omega (\tau )$.

To obtain the transformation properties of the plane waves, we use $(3.14)$ and
$(3.15)$ along with $(4.1)$ to get 
$$ 
{\bf E}'=\Re \left\{ {{\bf E}'_0\exp \left[ {i\left(
{{\bf k}\cdot {\bf x}-{1 \over c}\int_0^\tau {\omega (s)b(s)ds}} \right)} \right]}
\right\},
\eqno(4.3a) 
$$ 
$$ 
{\bf B}'=\Re \left\{ {{\bf B}'_0\exp \left[ {i\left( {{\bf
k}\cdot {\bf x}-{1 \over c}\int_0^\tau{\omega (s)b(s)ds}} \right)} \right]} \right\},
\eqno(4.3b) 
$$ 
with
$$ 
{\bf E}'_0=\gamma \left[ {{\bf E}_0+{1 \over c}\left( {{\bf v}\times {\bf B}_0}
\right)} \right]-(\gamma -1){{({\bf E}_0\cdot{\bf v})} \over {{\bf v}^2}}{\bf v}.
\eqno(4.3c) 
$$ 
$$ 
{\bf B}'_0=\gamma \left[ {{\bf B}_0-{1 \over c}\left( {{\bf
v}\times {\bf E}_0} \right)} \right]-(\gamma -1){{({\bf B}_0\cdot{\bf v})} \over
{{\bf v}^2}}{\bf v}. 
\eqno(4.3d) 
$$ 
We now use the inverse transformations $(2.1a)$, $(2.3b)$, and $(2.4)$ to transform
the phase
 $$
 \Phi =i\left( {{\bf k}\cdot {\bf x}-({1
\mathord{\left/ {\vphantom {1 c}} \right. \kern-\nulldelimiterspace} c})\int_0^\tau
{\omega(s)b(s)ds}} \right)
\eqno(4.3e) 
$$ 
in $(4.3a)$ and $(4.3b)$ to the corresponding expression in the primed variables: 
$$ 
\Phi' =i\left( {{\bf k}'\cdot {\bf
x}'-({1 \mathord{\left/ {\vphantom {1 c}} \right.\kern-\nulldelimiterspace}
c})\int_0^\tau {\omega' (s)b'(s)ds}} \right),
\eqno(4.3f) 
$$  
where the wave number ${\bf k}'$ and the frequency $\omega'(s)$ are to be
determined by the requirement that the transformed phase $\Phi'$ has the indicated
form.  Substituting $(2.1b)$ and
$(2.4)$ into $(4.3e)$, we get  
$$ 
\eqalign{&\Phi =i\left[ {\left( {{\bf k}+(\gamma
({\bf v})-1)\left( {{{{\bf k}\cdot {\bf v}} \over {||{\bf v}||^2}}} \right){\bf v}}
\right)\cdot {\bf x}'} \right.\cr  &\left. {\ \ \ \ \ \ \ \ \ \ \ \ +\gamma ({\bf v}){{{\bf
k}\cdot {\bf v}} \over c}\int\limits_0^\tau {b'(s)ds}-{1 \over c}\int_0^\tau  {\omega
(s)b'(s)ds}} \right]\cr
  &=i\left[ {\left( {{\bf k}+(\gamma -1) \left( {{{{\bf k}\cdot {\bf v}} \over {||{\bf
v}||^2}}} \right){\bf v}} \right)\cdot {\bf x}'} \right.\cr   &\left. {\ \ \ \ \ \ \ \ \ \
-{\gamma  \over c}\int\limits_0^\tau  {(\omega (s)-{\bf k}\cdot {\bf
v})b'(s)ds}-{\gamma  \over {c^2}}\int_0^\tau {\omega (s){\bf u}'\cdot {\bf v}ds}}
\right].\cr}  
\eqno(4.4) 
$$  
Integrating the last term in $(4.4)$ by parts, we obtain the
desired form for $\Phi' $, where the frequency relation is given by 
$$ 
\omega' (\tau
)=\gamma (\omega (\tau )-{\bf k}\cdot {\bf v}),\eqno(4.5) $$  and the wave number
relation (contributing the nonlocal part to $\Phi' $) is given by: $$ \eqalign{&{\bf
k}'\cdot {\bf x}'(\tau )={\bf k}\cdot {\bf x} '(\tau )+(\gamma -1)\left[{{{({\bf k}\cdot
{\bf v})({\bf v}\cdot {\bf x}'(\tau ))} \over {||{\bf v}||^2}}} \right]\cr  &-{{\gamma
\omega (\tau )} \over {c^2}}({\bf v}\cdot {\bf x}'(\tau ))+{{\gamma \omega (0)}
\over {c^2}}({\bf v}\cdot {\bf x}'(0))+{\gamma  \over {c^2}}\int_0^\tau 
{{{d\omega (s)} \over {ds}}\left[ {{\bf v}\cdot {\bf x}'(s)} \right]ds}.\cr}
\eqno(4.6)
$$

The wave vectors in our two frames differ by an extra nonlocal term compared to the
standard result, while the transformations of the frequencies $(4.5)$ agree with the
normal case except for the $\tau $ dependence.  This nonlocal term occurs because we
allowed the frequency of the wave to vary.  It is easy to check that, if $\omega $ is
constant (and the source passes though the origin(s) at $\tau =0$), we get the standard
result.

We now consider the planar representation $(4.6)$ with the velocity ${\bf v}$ taken
along the ${\bf x}={\bf x}'$ axes with angle $\theta $ defined as that between ${\bf
k}$ and ${\bf v}$, and $\theta'$ the angle between ${\bf k'}$ and ${\bf v}$,
$\omega$ constant, and assume that the source passes though the origin(s) at $\tau
=0$. We then obtain from $(4.6)$ the following relations between the angles $\theta$
and $\theta'$: 
$$ 
k'cos \theta' =\gamma k\cos \theta -\gamma {\omega  \over {c^2}}v,
\eqno{(4.7)} 
$$ 
$$ 
k'sin \theta' =k\sin \theta . \eqno{(4.8)} 
$$ 
They combine in the standard manner to give 
$$ 
\tan \theta' ={1 \over \gamma }{{\sin \theta } \over {\cos
\theta -{v \over c}{\omega  \over {kc}}}}. 
\eqno(4.9a) 
$$ 
This is the standard result for the aberration of wave vectors due to the relative motion
of the two reference frames.  It should be noted that we have not assumed that the $X$
frame is at rest relative to the medium.  Furthermore, we see from $(4.2a)$ that
$kc=\omega$ in free space (under the above assumptions).  In general, $kc=\omega
(\tau )\left[ {1+i\left( {{{c\dot \omega (\tau )} \mathord{\left/ {\vphantom {{c\dot
\omega (\tau )} {b\omega (\tau)^2}}} \right. \kern-\nulldelimiterspace} {b\omega (\tau )^2}}}
\right)} \right]^{1/2}$ so that our theory allows for nonlocal effects.

For any homogeneous medium, ${\omega  \mathord{\left/ {\vphantom {\omega  {\Re
k}}} \right. \kern- \nulldelimiterspace} {\Re k}}$ is equal to the phase velocity,
$v_{ph}$, of the wave, 
$$
 v_{ph}=c{\Re \left[ {1+i\left( {{{c\dot \omega (\tau )}
\mathord{\left/ {\vphantom {{c\dot \omega (\tau )} {b\omega (\tau )^2}}} \right.
\kern-\nulldelimiterspace} {b\omega (\tau )^2}}} \right)} \right]^{-1/2}}, 
\eqno(4.10)
$$ 
and ${c \mathord{\left/ {\vphantom {c {v_{ph}}}} \right.
\kern-\nulldelimiterspace} {v_{ph}}}$ is defined to be the index of refraction, $n$, of
the medium.  Thus, $(4.9a)$ becomes: 
$$
 \tan \theta' ={1 \over \gamma }{{\sin \theta
} \over {\cos \theta -{v \over {cn}}}}. \eqno(4.9b) 
$$  
This is what we would
normally expect from the standard theory.  However, the importance of $(4.10)$
becomes clear when we consider the group velocity, rather than the phase velocity, of
electromagnetic waves.  As is well-known, the group velocity represents the rate of
energy transmission, and is defined by $v_g=\Re({d\omega }/{dk})$.  We know that
use of observer clocks (proper-times) gives  $v_g=v'_g=c$.  The question is, what is
this relationship in the source proper-time theory ?

To determine how $v_g$ is related to $v'_g$, we restrict ourselves to the case when
the waves are moving parallel to the motion of the $X'$ frame relative to the $X$
frame, so that the wave vectors ${\Re {\bf k}}$ and ${\Re {\bf k}}'$ are parallel to
the velocity ${\bf v}$.  Then the frequency and wave number relations $(4.5)$ and
$(4.6)$ become (under these conditions)  
$$ 
\omega '(\tau )=\gamma (\omega (\tau
)-{\bf k}\cdot {\bf v}),  
\eqno(4.11)  
$$  
$$ 
k'x'=\gamma \left( {k-{{\gamma v\omega
(\tau )} \over {c^2}}} \right)x'(\tau )+{{\gamma v\omega (0)} \over
{c^2}}x'(0)+{{\gamma v} \over {c^2}}\int_0^\tau {{{d\omega (s)} \over {ds}}\left[
{x'(s)} \right]ds}, 
\eqno(4.12)  
$$ 
where, in the last equation, we have replaced the vector ${\bf x}'(\tau)$ by the scalar
$x'(\tau )$ because we are only interested in the $\tau$ dependence of the frequencies
and wave numbers.

Defining the group velocity in the $X, \, X'$ frames by  
$$ 
v_g\equiv \Re({{d\omega
} \over {dk}})=\Re ({{{{d\omega } \over {d\tau }}} \mathord{\left/ {\vphantom
{{{{d\omega }\over {d\tau }}} {{{dk} \over {d\tau }}}}} \right.
\kern-\nulldelimiterspace} {{{dk} \over {d\tau}}}}),\ \ \ \ \ v'_g\equiv \Re
({{d\omega '}\over{dk'}})=\Re({{{{d\omega '} \over {d\tau }}}\mathord{\left/
{\vphantom {{{{d\omega '} \over {d\tau }}} {{{dk'} \over {d\tau}}}}} \right.
\kern-\nulldelimiterspace} {{{dk'}\over {d\tau }}}}),
\eqno(4.13)  
$$   
we obtain from $(4.11)$ the equation  
$$ 
{{d\omega '} \over {d\tau }}=\gamma
\left({{{d\omega } \over {d\tau }}-v{{dk} \over {d\tau }}}\right),
\eqno(4.14)  
$$  
and from $(4.12)$ (after canceling terms),  
$$ 
{{dk'} \over {d\tau }}x'(\tau )=\gamma {{dk} \over {d\tau }}x'(\tau ). 
\eqno(4.15)   
$$  
Substitution of $(4.14)$ and $(4.15)$ into $(4.13)$ gives the relation 
$$ 
v_g=v'_g-v
\eqno(4.16) 
$$  
between the group velocities in the $X$ and $X'$ frames respectively.  It is clear that,
if the group velocity of the source has the value $c$ in one frame, it will not have that
value in the other frame and, indeed, may have a larger value.  Furthermore, the
Doppler formula $(4.11)$ can be written as  
$$
\omega'(\tau )=\gamma \omega (\tau )(1-\beta n\left[ {\omega (\tau )} \right]\cos \theta
),\eqno(4.17)  
$$
 where we have used $\beta ={v \mathord{\left/ {\vphantom {v c}} \right.\kern-
\nulldelimiterspace } c}$, $k={\omega  \mathord{\left/ {\vphantom {\omega 
{v_{ph}}}} \right.\kern-\nulldelimiterspace} {v_{ph}}}$,  and $n={c \mathord{\left/
{\vphantom {c {v_{ph}}}} \right. \kern-\nulldelimiterspace} {v_{ph}}}$.  Because
of $(4.10)$, this is a nonlinear relationship.

\bigskip 
\noindent {\bf 5.0 Particle Theory}

\medskip 
\noindent  {\bf 5.1 One-Particle Theory}

\bigskip 
\noindent  In order to understand the additional changes implied by fixing the
proper-time of the source for all observers, we need only consider the question of
particle dynamics.  Since our motivation is quantum theory, any change of variables
must be canonical.  (We focus on the $X$-frame equation, but the same results can
also be derived for the $X'$-frame.)  In the conventional formulation of quantum
theory, the Hamiltonian $H$ is the generator of observer proper-time translations.  We
now seek to identify the Hamiltonian $K$ which will generate source proper-time
translations.  To see how this may be done, let $W$ be any classical observable so that
the Poisson bracket defines Hamilton's equations in the $X$ frame by: (here, $H=\sqrt
{c^2{\bf p}^2+m^2c^4}$)  
$$ 
 {{dW} \over {dt}}={{\partial H} \over {\partial {\bf
p}}}{{\partial W} \over {\partial {\bf x}}}-{{\partial H} \over {\partial {\bf
x}}}{{\partial W} \over {\partial {\bf p}}} = \left\{ {H,W} \right\}.  
\eqno{(5.1)} 
$$ 
Now use the fact that the Hamiltonian for a free particle of mass $m$ can be
represented as $H=mc^2\gamma ({\bf w})$, so that $\gamma({\bf w})=H/mc^2$. 
This implies that    
$$  
d\tau=(mc^2/H)\,dt.$$  
The time evolution of  the functional W  is given by the chain rule:   

$$ {{dW} \over {d\tau }}={{dt} \over {d\tau }}{{dW}
\over {dt}}={H \over {mc^2}}\left\{ {H,W} \right\}. 
\eqno{(5.2)} 
$$

The energy functional $K$ conjugate to the proper-time $\tau$  must satisfy $\{K,W\}
=(H/mc^2) \{H,W\}$.  The  direct solution is obtained by rewriting the Poisson
bracket relation in (5.2) as  
$$ 
\eqalign{&{{dW} \over {d\tau }}=\left[ {{H \over
{mc^2}}{{\partial H} \over {\partial {\bf p}}}} \right]{ {\partial W} \over {\partial
{\bf x}}}-\left[ {{H \over {mc^2}}{{\partial H} \over {\partial {\bf x}}}}
\right]{{\partial W} \over {\partial {\bf p}}}\cr
  &\qquad ={\partial  \over {\partial {\bf p}}}\left[ {{{H^2} \over {2mc^2}}+a}
\right]{{\partial W} \over  {\partial {\bf x}}}-{\partial  \over {\partial {\bf x}}}\left[
{{{H^2} \over {2mc^2}}+a} \right]{{\partial W} \over {\partial {\bf p}}}.\cr} 
\eqno{(5.3)} 
$$

\noindent	 Now impose the condition that  ${\bf p}=0\Rightarrow K=H=mc^2$.  This
gives $a=a'=mc^2/2$, and   
$$ 
K={{H^2} \over {2mc^2}}+{{mc^2} \over 2}={{{\bf
p}^2} \over {2m}}+mc^2. 
\eqno{(5.4)} 
$$  
This equation was derived by Gill and Lindesay$^{48}$.  It looks like the
nonrelativistic case but is fully relativistic  and (partially) eliminates the problems
associated with the square root in the conventional implementation.  The most general
solution is  
$$ 
K=mc^2+\int_{mc^2}^H {(dt/d\tau )d\bar H} = mc^2+\int_{mc^2}^H
{(\bar H/mc^2)d\bar H}. 
\eqno{(5.5)} 
$$		 
There are three possible solutions to this equation depending on the assumptions
made.

{\bigskip \narrower \noindent {\bf  1.} If we fix the Lorentz frame, then $H/mc^2$ is
constant and we get \smallskip} 
$$ 
K={{H^2} \over {mc^2}} ={{{\bf p}^2} \over
{m}}+mc^2. 
\eqno{(5.6)} 
$$

{ \medskip \narrower \noindent			 This form was first derived by Gill$^{49}$, and
used to give a particle representation for the Klein-Gordon  equation with positive
probability density and with the source proper-time as an operator. \smallskip}

{ \bigskip \narrower \noindent {\bf 2.} If we keep the mass fixed and allow the
Lorentz frame to vary (boost), we get equation (5.4). \bigskip}

{ \smallskip \narrower \noindent {\bf 3.} If we keep the momentum ${\bf P}={\bf
P}_0$ fixed and allow the Lorentz frame $H$ and the mass $m$ to vary, we get
\smallskip} 
$$ 
K=mc^2=\sqrt {H^2-c^2{\bf P}_0^2}. 
\eqno{(5.7)} 
$$  

{ \medskip \narrower
\noindent								  This is the appropriate Hamiltonian in the constant momentum frame. 
This form has received the most attention,  having been used to associate the source
proper-time with the (off-shell) mass operator in parametrized relativistic  quantum
theories.  See Aparicio et al$^{50}$ for a recent discussion of this case.  The book by
Fanchi$^{51}$ surveys  all work up to 1993 (see also Fanchi$^{52}$). 	In all three
cases, a generator can be constructed proving that they are true canonical
transformations.  For the first two cases, the generators are constructed in references
48 and 49 respectively.  The construction  of the generator for the third case was done
in the seminal work of Bakamjian and Thomas $^{15}$. \bigskip}

We plan to use equation (5.4) in our work for a number of interesting reasons.  First, it
is simple, directly related to the nonrelativistic case, and the quantized version is (will
be) positive definite.  Furthermore, since the mass is fixed, it, along with the spin, are
natural choices to label the irreducible representations of the (proper-time) Poincar\' e
algebra describing elementary particles (see equations (5.24)-(5.32) and
Wigner$^{53}$).  In addition, it should be noted that some of the best models for
quark dynamics within nucleons ``appear" to be nonrelativistic (see, for example,
Strobel$^{54}$ and references therein).

The following theorem provides an explicit representation of the generator for the
canonical change of variables  for (5.4). (The result can be proved  by direct
computation$^{55}$.)

\medskip \noindent  {\bf Theorem 5.1.}  {\sl If $S=(mc^2-K)\tau,$ then $S$ is the
generator for the canonical change of variables from $ ({\bf x},{\bf p}, t, H)$ to
$({\bf x},{\bf p},\tau, K)$ (by our X-frame observer) and}: 
$$ 
{\bf p} \cdot d{\bf x}- Hdt={\bf p} \cdot d{\bf x} - Kd\tau  + dS. 
\eqno{(5.8)}
$$									 
{\sl It follows that the proper-time (free particle) equations will be form invariant
(covariant) for all observers}. 

\bigskip 
\noindent  {\bf 5.2 Many-Particle Theory}

\bigskip 
\noindent  Suppose we have a closed system of n particles with individual
Hamiltonians $H_i $ and total Hamiltonian $H$ (in the X-frame).  We assume that
$H$ is of the form 
$$ 
H = \sum\limits_{i = 1}^n {H_i }.  
\eqno{(5.9)} 
$$

\noindent If we define the effective mass $M$ and total momentum $ {\bf P}$ by
$$ 
Mc^2  = \sqrt {H^2  - c^2 {\bf P}^2 },\quad {\bf P} = \sum\limits_{i = 1}^n {{\bf
p}_i } , 
\eqno{(5.10)} 
$$          
$H$ also has the representation
$$ 
H = \sqrt {c^2 {\bf P}^2  + M^2 c^4 } . \eqno{(5.11)} $$			 
To construct the many-particle theory, we observe that the representation 
$$ d\tau  = (Mc^2 /H)dt \eqno{(5.12)} $$
						
\noindent does not depend on the number of particles in the system.  Thus, we can
uniquely define the  proper-time of the system for all observers. (In the primed frame,
we have a similar representation.)    If we let ${\bf L}$  be the boost (generator of pure
Lorentz transformations) and define the total angular momentum ${\bf J}$ by   
$$ 
{\bf J} = \sum\limits_{i = 1}^n {{\bf x}_i  \times {\bf p}_i } , 
\eqno{(5.13)}  
$$  
we then have the following Poisson Bracket relations characteristic of the algebra for
the Poincar\' e group (when we use  the observer proper-time):   
$$ 
{d{\bf  P}
\over{dt} }= \left\{ {H,{\bf P}} \right\} = {\bf 0}  \qquad  {d{\bf J} \over {dt}} =
\left\{ {H,{\bf J}} \right\} = {\bf 0}  \qquad  \left\{ {P_i ,P_j } \right\} = 0  
\eqno{(5.14)}   
$$  
$$ 
{
\left\{ {J_i ,P_j } \right\} = \varepsilon _{ijk} P_k  }  \qquad  { \left\{ {J_i ,J_j }
\right\} = \varepsilon _{ijk} J_k } \qquad  { \left\{ {J_i ,L_j } \right\} = \varepsilon
_{ijk} L_k }    
\eqno (5.15)  
$$  
$$ 
{ {d{\bf  L}\over dt}
 = {\left\{ H,{\bf L} \right\} } = - {\bf P}  }  \qquad  { \left\{ {P_i , L_j } \right\} =  -{
\delta _{ij}} H/c^2, } \qquad  { \left\{ {L_i ,L_j } \right\} =  - {\varepsilon _{ijk}}
J_k /c^2. }  
\eqno (5.16) 
$$  
It is easy to see that $M$ commutes with $H$, ${\bf P}$,  and ${\bf J}$, and to show
that $M$ commutes with ${\bf L}$. Constructing $K$ as in the one-particle case, we
have  
$$ 
K = {{H^2 } \over {2Mc^2 }} + {{Mc^2 } \over 2} = {{{\bf P}^2 } \over
{2M}} + Mc^2 . 
$$					 
Thus, we can use the same definitions for ${\bf P}$, ${\bf J}$, and ${\bf L}$  to
obtain our new commutation relations: 
$$ 
{{d{\bf P}} \over {d\tau }} = \left\{ {K,{\bf P}} \right\} = {\bf 0}, \quad {{d{\bf
J}} \over {d\tau }} = \left\{ {K,{\bf J}} \right\} = {\bf 0}, \quad \left\{ {P_i ,P_j }
\right\} = 0,
 \eqno{(5.17)} 
$$ 
$$ 
\left\{ {J_i ,P_j } \right\} = \varepsilon _{ijk} P_k, 
\quad
\left\{ {J_i ,J_j } \right\} = \varepsilon _{ijk} J_k , \quad \left\{ {J_i ,L_j } \right\} =
\varepsilon _{ijk} L_k, 
\eqno{(5.18)}  
$$ 
$$ 
{{d{\bf L}} \over {d\tau }} = \left\{
{K,{\bf L}} \right\} = {{ - H} \over {Mc^2 }}{\bf P}, \quad \left\{ {P_i ,L_j } \right\}
=  - \delta _{ij} H/c^2,  \quad \left\{ {L_i ,L_j } \right\} =  - \varepsilon _{ijk} J_k
{\kern 1pt} /{\kern 1pt} c^2. 
\eqno{(5.19)}  
$$			 
It follows that, except for a constant
scale change, the proper-time group is generated by the same algebra as the  Lorentz
group. This result is not surprising given the close relation between the two groups.  It
also proves our earlier  statement that the form of $K$ is fully relativistic.

Let the map from $( {\bf x}_i, t)\, \to \,({\bf x}_i, \tau)$ be denoted by ${\bf
C}[\,t,\,\tau]$, and let  ${\bf P}(X', X)$ be the Poincar\' e map from $X \to X'$.
  
\medskip
 \noindent  {\bf Theorem 5.2} {\sl The proper-time coordinates of the system
as seen by an observer at $X$ are related to those of an observer  at $X'$ by the
transformation}: 
$$ 
{\bf R}_M[\tau ]={\bf C}[ \,t',\,\tau ]{\bf P}(X',\,X){\bf C}^{-1}[ \,t,\,\tau ]. 
\eqno{(5.20)} 
$$

\medskip 
\noindent  {\bf  Proof:}  The proof follows since the diagram below is
commutative.   
$$
 \matrix {X(\{{\bf x}_i \},\, t)  &  {\rm  {}} & { \longrightarrow} 
&  {\rm  {}} & X'(\{{\bf x'}_i \},\, t') \cr &  {\rm  {}}&  {\rm  {}}&  {\rm  {}}& 
{\rm  {}}&  {\rm  {}}\cr &  {\rm  {}}&  {\rm  {}}&  {\rm  {}}&  {\rm  {}}&  {\rm 
{}}\cr {{{\bf C}^{-1}[\,t,\,\tau]}} &   \Bigg\uparrow &   {\rm  {}} & 
\Bigg\downarrow & {{\bf C}[\,t',\,\tau]}\cr &  {\rm  {}}&  {\rm  {}}&  {\rm  {}}& 
{\rm  {}}&  {\rm  {}}\cr &  {\rm  {}}&  {\rm  {}}&  {\rm  {}}&  {\rm  {}}&  {\rm 
{}}\cr X(\{{\bf x}_i \},\, \tau)  &  {\rm  {}} & \longleftarrow &  {\rm  {}} &
X'(\{{\bf x'}_i \},\, \tau)\cr} 
 \eqno{(5.21)} 
$$

The top diagram is the Poincar\' e map from $X \to X'$.  {\sl It is important to note that
this map is between  the coordinates of observers}.  In this sense, our approach may be
viewed as a direct generalization of the conventional theory.  In the global case, when
${\bf U}$ is constant, $t$ is related to $\tau$  by a scale transformation so that we
have a group with the same algebra as the Poincar\' e group (up to a constant scale),
but it has an Euclidean metric!  In this case, Theorem 5.2 proves that ${\bf R}_M$ is
in the proper-time group, formed by a similarity action on  the Poincar\' e group by the
canonical group ${\bf C}_\tau$.    On the other hand, Theorem 5.2 is true in general.
This means that in both the local and global cases (when the acceleration is nonzero)
$t$ is related to $\tau_i$ and $\tau$ via nonlocal (nonlinear) transformations.  It
follows that, in general, the group action is not linear, and hence is not covered by the
Cartan classification.  
 
Since $K$ does not depend on the center-of-mass position ${\bf X}$, it is easy to see
that  
$$ 
{\bf U} = {{d{\bf X}} \over {d\tau }} = {{\partial K} \over {\partial {\bf
P}}} = {{\bf P} \over M} =  {1 \over M}\sum\limits_{i = 1}^n {m_i {\bf u}_i },
\eqno{(5.22)}  
$$ 
where ${\bf u}_i  = d{\bf x}_i /d\tau _i $.  We can now define $b$ by 
$$ 
b = \sqrt {{\bf U}^2  + c^2 }  \Rightarrow H = Mcb. \eqno{(5.23)} 
$$ 
Thus, equation $(5.12)$ can also be represented as 
$$ 
d\tau  = (c/b)dt. \eqno{(5.24)} $$

If we set ${\bf v}_i  = d{\bf x}_i /d\tau $, an easy calculation shows that   
$$ {\bf
u}_i  = {{d{\bf x}_i } \over {d\tau _i }} = {{d\tau } \over {d\tau _i }}{{d{\bf x}_i }
\over {d\tau }} =  {{b_i } \over b}{\bf v}_i  \Rightarrow {{{\bf u}_i } \over {b_i }} =
{{{\bf v}_i } \over b}.  
\eqno{(5.25)} 
$$

The velocity ${\bf v}_i $ is the one our observer sees when he uses the global
proper-clock of the system to compute  the particle velocity, while ${\bf u}_i $ is the
one seen when he uses the local proper clock of the particle to compute its velocity. 
Solving for ${\bf u}_i $ and $b_i $ in terms of ${\bf v}_i $ and $b$, we get  
$$ {\bf
u}_i  = {{c{\bf v}_i } \over {\sqrt {b^2  - \mathop {\bf v}\nolimits_i^2 }}}, {\rm  
}b_i  =  {{cb} \over {\sqrt {b^2  - \mathop {\bf v}\nolimits_i^2 }}} \,\,\,  {\rm or }
\,\,\,  {{b_i } \over b} = {c \over {\sqrt {b^2  - \mathop {\bf v}\nolimits_i^2 } }}.
\eqno{(5.26)}  
$$

Note that, since $b^2  = {\bf U}^2  + c^2$, if ${\bf U}$ is not zero, then any ${\bf
v}_i $ can be larger than $c$.   On the other hand, if  ${\bf U}$ is zero, $b = c$ and,
from the global perspective, our theory looks like the  conventional one. Using (5.26),
we can rewrite ${\bf U}$ as  
$$ 
{\bf U} = {1 \over M}\sum\limits_{i = 1}^n {m_i
{\bf u}_i }  =  {1 \over M}\sum\limits_{i = 1}^n {{{m_i c{\bf v}_i } \over {\sqrt
{b^2  - \mathop {\bf v}\nolimits_i^2 } }}}  = {1 \over M}\sum\limits_{i = 1}^n
{{{b_i m_i {\bf v}_i } \over b}}  = {1 \over H}\sum\limits_{i = 1}^n {H_i {\bf v}_i
}.   \eqno{(5.27)} 
$$  
It follows that the position of the center-of-mass (energy) satisfies   
$$ 
{\bf X} = {1 \over H}\sum\limits_{i = 1}^n {H_i {\bf x}_i }  + {\bf Y},
\quad {d{\bf Y} \over d{\tau}}={\bf 0}. 
\eqno{(5.28)}  
$$ 
It is natural to choose ${\bf Y}$ so that ${\bf X}$ is the canonical center of mass:   
$$
{\bf X} = {1 \over H}\sum\limits_{i = 1}^n {H_i {\bf x}_i }  +  {{c^2 ({\bf S} \times
{\bf P})} \over {H(Mc^2  + H)}}, 
\eqno{(5.29)} 
$$		              
where  ${\bf S}$ is the (conserved) spin of the system.  The important point is that 
${\rm (}{\bf X},{\bf P},\tau, K{\rm )}$ is the new set of (global) variables for the
system.  

\medskip  
\noindent
 {\bf Theorem 5.3} {\sl If $S = {\rm (}mc^{2} -K{\rm )}\tau$, then $S$ is the
generator for the change of variables from  $ \left( {\{ {\bf x}_i \} {\rm ,\{ }{\bf p}_i
\} {\rm ,t, }H} \right) \to {\rm (}\{ {\bf x}_i \} {\rm ,\{ }{\bf p}_i \} {\rm ,}\tau {\rm ,
}K{\rm )}$, from  $ {\rm (}{\bf X}{\rm ,}{\bf P}{\rm ,t, }H{\rm )} \to {\rm (}{\bf
X}{\rm ,}{\bf P},\tau {\rm , }K{\rm )} $, and}:  
$$ 
\sum\limits_{i = 1}^n {{\bf p}_i
d{\bf x}_i }  - Hdt=\sum\limits_{i = 1}^n {{\bf p}_i d{\bf x}_i }  - Kd\tau  + dS,
\eqno(5.30) 
$$ 
$$ 
{\bf P} \cdot d{\bf X} - Hdt{\rm   =  }{\bf P} \cdot d{\bf X} -
Kd\tau  + dS.
 \eqno(5.31) $$	

We can now write down the transformations that fix the proper-time of the system of
particles for any observer.  If $ {\bf V}$ is the relative velocity between two
observers, we have  
$$ 
b' = \gamma ({\bf  V})\left[ {b - {\raise0.7ex\hbox{${{\bf  U}
\cdot {\bf  V}}$} \!\mathord{\left/
 {\vphantom {{{\bf  U} \cdot {\bf  V}} c}}\right.\kern-\nulldelimiterspace}
\!\lower0.7ex\hbox{$c$}}} \right],\;{\rm     }b = \gamma ({\bf V})\left[ {b' + 
{\raise0.7ex\hbox{${{\bf U}' \cdot {\bf  V}}$} \!\mathord{\left/
 {\vphantom {{{\bf  U}' \cdot {\bf  V}} c}}\right.\kern-\nulldelimiterspace}
\!\lower0.7ex\hbox{$c$}}} \right], 
\eqno{(5.32)} 
$$ 
$$ 
{\bf  X}' = \gamma ({\bf
V})\left[ {{\bf  X}^{\bf  \dag }  - {{({\bf  V}} \mathord{\left/
 {\vphantom {{({\bf V}} c}} \right.
 \kern-\nulldelimiterspace} c})b\tau } \right],\,\,\,\,\,\,\,\,{\bf  X} = \gamma ({\bf 
V})\left[ {{\bf  X}'^{\bf  \dag }  +  {{({\bf  V}} \mathord{\left/
 {\vphantom {{({\bf  V}} c}} \right.
 \kern-\nulldelimiterspace} c})b'\tau } \right], \eqno{(5.33)} $$ $$ {\bf U}' = \gamma
({\bf V})\left[ {{\bf U}^{\bf \dag }  - ({{\bf V} \mathord{\left/
 {\vphantom {{\bf V} c}} \right.
 \kern-\nulldelimiterspace} c})b} \right],\,\,\,\,\,\,\,{\bf U} = \gamma ({\bf V})\left[
{{\bf U}'^{\bf \dag }  +  ({{\bf V} \mathord{\left/
 {\vphantom {{\bf V} c}} \right.
 \kern-\nulldelimiterspace} c})b'} \right].  
\eqno{(5.34)}  
$$  
As our system is closed, ${\bf U}$ is constant and $\tau $ is linearly related to $t$.
Yet, the physical interpretation is different in the extreme if ${\bf U}$ is not zero. 
Furthermore, we see from  equation (5.34) that, even if ${\bf U}$ is zero in one frame,
it will not be zero in any other frame that is in relative motion.  It is clear that $\tau $
is uniquely determined by the particles in the system and is available to all observers. 
Just as  important is the fact that there is a very basic relationship between the global
system clock and the clocks of the individual particles.  In order to derive this
relationship, we return to our definition of the global Hamiltonian $K$ and let $W$ be
any observable.  Then  
$$ 
\eqalign{
  & {{dW} \over {d\tau }} = \left\{ {K,W} \right\} = {H \over {Mc^2 }}\left\{ {H,W}
\right\} =  {H \over {Mc^2 }}\sum\limits_{i = 1}^n {\left\{ {H_i ,W} \right\}}   \cr 
  & \qquad = {H \over {Mc^2 }}\sum\limits_{i = 1}^n {{{m_i c^2 } \over {H_i
}}\left[ {{{H_i } \over {m_i c^2 } }\left\{ {H_i ,W} \right\}} \right]}  =
\sum\limits_{i = 1}^n {{{Hm_i } \over {MH_i }}\left\{ {K_i ,W} \right\}}  \cr}. 
\eqno{(5.35)} 
$$ 	  
Using the (easily derived) fact that $d\tau _i /d\tau  = Hm_i
/MH_i  = b_i /b$, we get 		  
$$ 
{{dW} \over {d\tau }} = \sum\limits_{i = 1}^n
{{{d\tau _i } \over {d\tau }}\left\{ {K_i ,W} \right\}}. 
\eqno{(5.36)}  
$$            
Equation (5.36) is very important because it relates the global systems dynamics to the
local system's dynamics and  provides the basis for a direct approach to the quantum
relativistic many-body problem using one (universal) wave function.  The use of a
many-times approach is not new and dates back to the early work of Dirac et
al$^{56}$.  Our many-times approach is like that of Rohrlich and Horwitz$^{57}$
(see also Longhi et al$^{58}$).  Our approach is distinct, as is clear from (5.36) and
the fact that all our times are unique and invariant for all observers.

\bigskip 
\noindent  {\bf 5.3 Interaction (Global External)}

\bigskip 
\noindent  In this section, we follow convention (in the simplest fashion) and
introduce an external global interaction via minimal coupling in the free Hamiltonian. 
This means that we fix the position X, the momentum P, and the mass M.  It is still
possible for the angular momentum J to be conserved but, in general, it need not be
equal to the  angular momentum in the noninteracting case. Our interaction
Hamiltonian becomes  
$$ 
K = {{\Pi ^2 } \over {2M}} + Mc^2  + V({\bf X}),
\eqno(5.37) 
$$                 
where ${\bf A} = {\bf A}(X,\tau ),{\rm  }V = V(X,\tau)$ are
the vector and scalar potentials of the external field, and $ \Pi  = {\bf P} - ({\rm
q/c)}{\bf A}$.  (In the next section, we derive an alternative equation appropriate
when the cause of the external field is included in the theory to form a closed system.) 
Using (5.37) and Hamilton's equations, we get 
$$ 
{\bf \dot X} = {\bf U} = {\Pi  \over
{\rm M}},\,\,\,\,{\bf \dot P} =  - {{\nabla \Pi ^2 } \over {2{\rm M}}}\, - \nabla V.
\eqno{(5.38)} 
$$              
Using standard vector identities, elementary calculations
give the (proper-time) Lorentz force 
$$ 
{{Mc} \over b}{{d{\bf U}} \over {d\tau }} =
q{\bf E} + {q \over b}{\bf U} \times {\bf B}, 
\eqno{(5.39)} 
$$ 
$$ 
{\bf E} =  - {1
\over b}{{\partial {\bf A}} \over {\partial \tau }} - \nabla V,\,\,\,\,{\bf B} = \nabla 
\times {\bf A}.
 \eqno{(5.40)} 
$$

The fact that we can derive (a generalized form of) the Lorentz force from a
(apparently) nonrelativistic Hamiltonian  is well-known (see Hughes$^{59}$). 
However, in order to see how the nonuniqueness of the Maxwell-Lorentz theory
shows up here, we need only recall that ${\bf W}/c = {\bf U}/b$ and $\,(1/b)\partial
/\partial \tau  = (1/c)\partial /\partial t$, so we can also write equations (5.39) and
(5.40) as (${\bf W} = d{\bf X}/dt$)  $$ M{{d{\bf U}} \over {dt}} = q{\bf E} + {q
\over c}{\bf W} \times {\bf B}, \eqno{(5.41)}  $$  $$ {\bf E} =  - {1 \over
c}{{\partial {\bf A}} \over {\partial t}} - \nabla V,\,\,\,\,{\bf B} = \nabla  \times {\bf
A}. \eqno{(5.42)} $$  This is the "original" force derived by Lorentz$^3$ (in 1892)
and used as a part of his theory of the electrodynamics and optics of macroscopic
phenomena.   What is truly remarkable is the fact that the two equations (5.39) and
(5.41) are mathematically equivalent, but clearly not physically equivalent, with
radically different physical interpretations. 

\bigskip

\noindent{\bf Global Field Theory}

\medskip  \noindent We can now discuss the fields of our global system of particles in
a given external field.  Using $({{\rm 1} \mathord{\left/{\vphantom {{\rm 1} c}}
\right. \kern- \nulldelimiterspace} c})(\partial {\kern 1pt} {\kern 1pt} /{\kern 1pt}
\partial t) = ({{\rm 1} \mathord{\left/ {\vphantom {{\rm 1} b}} \right.    
\kern-\nulldelimiterspace} b})(\partial {\kern 1pt} /{\kern 1pt} \partial \tau )$  (as in
the one-particle case), we can write Maxwell's equations for the global system of
particles as:
$$ 
\nabla \cdot {\bf B}=0,\ \ \ \ \ \ \ \ \ \nabla \times {\bf E}+{1 \over b}{{\partial {\bf
B}} \over {\partial \tau }}=0,
\eqno (5.43a) 
$$
$$ 
\nabla \cdot {\bf E}=4\pi \rho ,\ \ \ \ \nabla \times {\bf B}={1 \over b}\left[
{{{\partial {\bf E}} \over {\partial \tau }}+4\pi {\bf J}} \right],
\eqno (5.43b)  
$$
where  ${\rho}$  and  ${\bf J}$ represent the charge  and current density of the system
(as a whole) relative to  its external  environment.  Taking the  curl  of  the  last 
equations of (5.43a) and (5.43b), using the standard vector identity (for any
sufficiently differentiable {\bf W})
$$ 
{ \nabla  \times (\nabla  \times {\bf W}) = \nabla (\nabla  \cdot {\bf W}) - \nabla ^2
{\bf W}}, 
$$ 
and the first equations of (5.43a) and (5.43b), we get the corresponding
global wave equations
$$
          \eqalign{
             &  {1  \over {\rm b}}{\partial  \over {\partial
     \tau  }}\left[  {{1 \over {\rm b}}{{\partial  {\bf  E}}
     \over  {\partial  \tau }}} \right] - \nabla  ^2   \cdot
     {\bf  E} =  - \nabla ({\rm 4}\pi \rho ) - {1 \over {\rm
     b}}{\partial   \over  {\partial \tau  }}\left[  {{{4\pi
     {\bf J}} \over {\rm b}}} \right],  \cr
             &  {1  \over {\rm b}}{\partial  \over {\partial
     \tau  }}\left[  {{1 \over {\rm b}}{{\partial  {\bf  B}}
     \over  {\partial  \tau }}} \right] - \nabla  ^2   \cdot
     {\bf  B}  = {1 \over {\rm b}}{\partial  \over {\partial
     \tau  }}\left[  {{{4\pi \nabla  \times {\bf  J}}  \over
     {\rm b}}} \right]. \cr}
 \eqno (5.44)  
$$

\noindent Computing  the  derivatives, these  equations  may  also  be written as
$$
 \eqalign{&{1  \over  {b^2}}{{\partial ^2{\bf  E}}  \over  {\partial \tau ^2}}-\left[
{{{\bf  U} \over {b^4}}\cdot {{d{\bf  U}} \over {\partial \tau  }}} \right]\left[
{{{\partial {\bf  E}} \over {\partial  \tau }}}   \right]-\nabla  ^2{\bf  E}=-\nabla  (4\pi 
\rho  )-{1   \over b}{\partial  \over {\partial \tau }}\left[ {{{4\pi {\bf J}}  \over b}}
\right],\cr
  &{1 \over {b^2}}{{\partial ^2{\bf  B}} \over {\partial \tau ^2}}- \left[ {{{\bf  U}
\over {b^4}}\cdot {{d{\bf  U}} \over {\partial \tau  }}} \right]\left[  {{{\partial  {\bf 
B}}  \over  {\partial  \tau   }}} \right]-\nabla  ^2{\bf  B}={1  \over b}{\partial   \over 
{\partial \tau   }}\left[   {{{4\pi  \nabla  \times  {\bf  J}}   \over   b}} \right].\cr} 
\eqno{(5.45)} 
$$

From (5.45), we see directly that the dissipative term does not depend on the gauge.
These equations imply that the field of the global system dissipates energy (radiation)
throughout the enclosing domain.  Since ${\bf U}{\rm  = (1}{\kern 1pt} {\rm /}{\kern
1pt} {\rm M)}\sum\nolimits_{i = 1}^n {{\rm m}_{\rm i} {\bf u}_{\rm i} } $, this
radiation depends on the average of the (local proper) motion of all the particles in the
system (e.g., ${\bf u}_i  = {{d{\bf x}_i } \mathord{\left/  {\vphantom {{d{\bf x}_i }
{d\tau _i }}} \right.  \kern-\nulldelimiterspace} {d\tau _i }}$).  This  suggests that 
the particles live in a heat bath of radiation created by the  global system's (inertial)
reaction to the external field.  This heat bath will fill out any domain enclosing the
system of particles.

When ${\bf U}$ is constant, ${\bf \dot U}={\bf 0}$ so that there are only velocity
fields (and no radiation fields).  This is necessarily the case if energy is conserved on
the global level and implies the following theorem:

\smallskip 
\noindent {\bf Theorem 5.4}  {\sl If ${\bf U}$ is constant then all radiation
generated by internal interactions must be absorbed by the particles in the system}.

The above theorem was a (required) postulate for the Wheeler-Feynman formulation. 
It should be noted that our formulation does not require advanced fields.  As will be
seen in the next Section, the individual particle interaction from the local point of view
(using the particle proper-time), is of the local field type.   In Section 5.5, we will see
that the individual particle interaction, from the global point of view (using the global
proper-time), is  of the action-at-a-distance type.  This confirms and refines the
Wheeler-Feynman conjecture concerning the relationship between these two views.

It is clear that, in general, the above theorem is only approximately true and it is more
reasonable to consider conservation of energy in a statistical sense.  For example, our
galaxy is clearly not a conserved system in the absolute sense, but may be considered
conserved in the mean. Thus, the radiation we receive from the other galaxies is, on
the average, equal to the radiation leakage from our galaxy.
 
\bigskip 
\noindent  {\bf 5.4 Interaction (Internal)}

\bigskip 
\noindent  In this section we assume that the system of n interacting particles
can be represented via: 
$$
          \eqalign{
             &  H = \sum\limits_{i = 1}^n {H_i }  =  H_0 + V,\quad H_{i}  =  H_{oi} +
V_{i} ,  \cr
             &  H_{0i}  = \sqrt { {c}^{2} \pi _i^2  + m_i^2 c^4 } ,\quad \pi _i  = {\bf p}_i 
- {{e_i } \over c}{\bf A}_i ,
 \cr} 
\eqno (5.46)  
$$
$$
          \eqalign{
             &  H_0  =  \sum\limits_{i = 1}^n {H_{0i} } ,\quad {\bf A}_i = \sum\limits_{i
\ne j} {{\bf A}_{ji} },\quad e_i {\bf A}_{ji}  = {{e_i e_j \left( {{\bf w}_j  - {\bf w}_i
} \right)} \over 2{s_{ji} }},  \cr
             &  V = \sum\limits_{i = 1}^n {V_i } ,\quad V_i  = \sum\limits_{i \ne j}
{{{e_i e_j } \over 2{s_{ij} }}},\quad  s_{ji}  = s_{ij} , \quad {\partial  \over {\partial
{\bf x}_i }}(s_{ij} ) =  - {\partial  \over {\partial {\bf x}_j }}(s_{ij} ).  \cr}
 \eqno (5.47)  
$$

Since we have specified the internal interactions, it is not a priori clear that the system
is closed.   Under the stated conditions, the following results can be proven by direct
computation.  

\medskip 
\noindent  {\bf Lemma 5.1} {\sl Set $ {\bf P} = \sum\limits_{i = 1}^n {{\bf
p}_i } ,\;\;\Pi  = \sum\limits_{i = 1}^n {\pi _i }  $, then $\Pi  = {\bf P}$.}

\bigskip \noindent  {\bf Theorem 5.5 } 
$ \left\{ {H,{\bf P}} \right\} = 0, \quad \left\{ {H,V} \right\} = 0, \quad \left\{ {{\bf
P},V} \right\} = 0. $

\medskip 
\noindent  It follows that, as in Section 5.2, we can define the total effective
mass $M$ by $Mc^2  = \sqrt {H_{}^2  - c^2 {\bf P}^2 }$, so that $ H = \sqrt {c^2
{\bf P}^2  + M^2 c^4 }$.

\bigskip 
\noindent  {\bf Lemma 5.2}  
$ \left\{ {H,M} \right\} = 0,\quad \left\{ {{\bf P},M} \right\} = 0. $

\medskip 
\noindent  Using the above results, it now follows that the set $\left\{ {\left.
{H_i } \right|}\;{1\le i \le n} \right\}$, forms a closed system satisfying all the
conditions of Section 5.2.  

\bigskip 
\noindent  {\bf 5.5 Particle Interaction (Local View)}

\bigskip  
\noindent  We are now ready to investigate the nature of the dynamics of the
ith-particle (say) caused by the action of the  other particles on it.  Since there are two
possible clocks, $\tau $ and $\tau _i $, there are two different views, or answers, to
our question.  Let $W_i $ be any observable of the ith-particle, then
$$
   \eqalign{
             &  {{dW_i } \over {d\tau }} = \left\{ {K,W_i } \right\} = \sum\limits_{j =
1}^n {{{\partial K} \over {\partial {\bf p}_j } }{{\partial W_i } \over {\partial {\bf
x}_j }} - {{\partial K} \over {\partial {\bf x}_j }}{{\partial W_i } \over {\partial {\bf
p}_j }}},  \cr
             &  {{dW_i } \over {d\tau _i }} = \left\{ {K_i ,W_i } \right\} =  {{\partial K_i
} \over {\partial {\bf p}_i }}{{\partial W_i } \over {\partial {\bf x}_i }} - {{\partial
K_i } \over {\partial {\bf x}_i }}{{\partial W_i } \over {\partial {\bf p}_i }},\cr}
\eqno (5.48)  
$$ 
$$ 
K = {{H^2 } \over {2Mc^2 }} + {{Mc^2 } \over 2},\quad K_i  = {{H_i^2 } \over
{2m_i c^2 }} + {{m_i c^2 } \over 2}. 
\eqno{(5.49)}  
$$    
The equations of motion can be computed rather easily in the second case.  The
Hamiltonian has an explicit representation as 
$$ 
K_i  = {{\pi _i^2 } \over {2m_i }} +
m_i c^2  + {{V_i^2 } \over {2m_i c^2 }} + {{H_{i0} V_i } \over {m_i c^2 }}, 
$$  
$$
\Rightarrow \, {{d{\bf x}_i } \over {d\tau _i }} = {{\partial K_i } \over {\partial {\bf
p}_i }} =  {{\pi _i^{} }
\over {m_i }}\left( {{{H_i } \over {H_{i0} }}} \right), 
\eqno{(5.50)} 
$$ 
$$ 
{{d{\bf
p}_i } \over {d\tau _i }} =  - {{\partial K_i } \over {\partial {\bf x}_i }} =   - {{\nabla
_i \pi _i^2 } \over {2m_i }}\left( {{{H_i } \over {H_{i0} }}} \right) - \nabla _i V_i
\left( {{{H_i } \over {m_i c^2 }}} \right). 
\eqno{(5.51)}  
$$

\noindent Using $ (H_i /H_{i0} )\nabla _i \pi _i^2  =  - 2(e_i /c)\left[ {({\bf u}_i  \cdot
\nabla _i ) {\bf A}_i  + {\bf u}_i  \times (\nabla _i  \times {\bf A}_i )} \right], \, {\bf
B}_i  = (\nabla _i  \times {\bf A}_i ), $

\noindent 
and $ (H_i /m_i c^2 ) = (b_i /c), $ we have 
$$ 
{{d{\bf p}_i } \over {d\tau _i }} =  {{e_i } \over c}\left[ {({\bf u}_i  \cdot \nabla
_i ){\bf A}_i  + {\bf u}_i  \times {\bf B}_i } \right] - {{b_i } \over c}\nabla _i V_i .
\eqno{(5.52)}
 $$ 

\noindent Finally, using $ ({\bf u}_i  \cdot \nabla _i ){\bf A}_i  = (d{\bf A}_i /d\tau _i
) - (\partial {\bf A}_i /\partial \tau _i ), $ and  $ V_i  = e_i \Phi _i , $ we have  
$$ 
{c \over {b_i }}{{d{\pi} _i } \over {d\tau _i }} =  e_i {\bf E}_i  + {{e_i } \over {b_i
}}\left( {{\bf u}_i  \times {\bf B}_i } \right), 
\eqno{(5.53)} 
$$ 	 
$$ 
{\bf E}_i  =  - {1
\over {b_i }}{{\partial {\bf A}_i } \over {\partial \tau _i }} - \nabla _i \Phi _i .
\eqno{(5.54)} 
$$ 

\noindent We call this the local view since it gives information about the action of the
external  field on the particle but provides no information about the action of the
particle on the source of the external force.  Equation (5.53) is of the same form as
(5.39), so if we use $ (1/b_i )(\partial /\partial \tau _i ) = (1/c)(\partial /\partial t)$ and $
({\bf u}_i /b_i ) = ({\bf w}_i /c)$, we have
$$
 {{d\pi _i } \over {dt}} = e_i {\bf E}_i  + {{e_i } \over c}\left( {{\bf w}_i  \times
{\bf B}_i } \right),
 \eqno{(5.55)} 
$$  
$$ 
{\bf E}_i  =  - {1 \over c}{{\partial {\bf A}_i
} \over {\partial t}} - \nabla _i \Phi _i . 
\eqno{(5.56)} 
$$

This is the same result we found in Section 5.3 when we used minimal coupling
directly for the global case.   	For later reference we return to equation (5.50), solve for
$\pi _i $, and differentiate, to get   
$$ 
{\bf \dot \pi} _i  = \bar m_i {\bf \dot u}_i  - \bar
m_i {\bf u}_i \left[  {{{({\bf u}_i  \cdot \nabla _i )V_i } \over {H_i }}} \right],\quad
\bar m_i  =  m_i \left( {1 - {{V_i } \over {H_i }}} \right). 
\eqno{(5.57)}
 $$    
Putting this term in (5.53), and taking the dot product, we have   $$ \left( {{\bf
u}_i  \cdot {\bf \dot u}_i } \right) =  {1 \over 2}{d \over {d\tau _i }}\left\| {{\bf u}_i }
\right\|^2  = \left\| {{\bf u}_i } \right\|^2 \left[ {{{({\bf u}_i  \cdot \nabla _i )V_i }
\over {H_i }}} \right] + {{e_i } \over {\hat m_i }}\left( {{\bf u}_i  \cdot {\bf E}_i },
\right).  
\eqno{(5.58)}  
$$  
$$ 
\hat m_i  = {c \over {b_i }}\bar m_i  = m_i {c \over
{b_i }}\left[ {1 - {{V_i } \over {H_i }}} \right]. 
$$  

\bigskip 
\noindent  {\bf 5.6 Particle Interaction (Global View)}

\bigskip 
\noindent  Let us now see what changes occur when we focus on the motion
of the same particle as seen from the global point of view.   In this case, we have   
$$
{{d{\bf x}_i } \over {d\tau }} = {\bf v}_i  = {{\partial K} \over {\partial {\bf p}_i }} =
 \left( {{H \over M}} \right){{\pi _i^{} } \over {H_{i0} }},  
\eqno{(5.59)}  
$$  
$$ 
{{d{\bf p}_i } \over {d\tau }} =  - {{\partial K} \over {\partial {\bf x}_i }} = 
 - \left( {{H \over {Mc^2 }}} \right)\sum\limits_{k = 1}^n  {\left[ {{{c^2 \nabla _i \pi
_k^2 } \over {H_{k0} }} - \nabla _i V_k } \right]}.  
\eqno{(5.60)} 
$$  
Using standard
calculations as in the local view, and $(H/Mc^2 ) = (b/c)$, we have  
$$ 
{{d{\bf p}_i }
\over {d\tau }} =  \sum\limits_{k = 1}^n {\left\{ {{{e_k } \over c}\left[ {({\bf v}_k 
\cdot \nabla _i ){\bf A}_k  + {\bf v}_k  \times (\nabla _i  \times {\bf A}_k )} \right] -
{b \over c}\nabla _i V_k } \right\}}.    
\eqno{(5.61)} 
$$  
Now use  
$$
 ({\bf v}_i  \cdot \nabla _i ){\bf A}_i  = (d{\bf A}_i /d\tau ) - (\partial
{\bf A}_i /\partial \tau ),  
$$
 to get    
$$ 
\eqalign{
  & {{d{\bf p}_i } \over {d\tau }} - {{e_i } \over c}{{d{\bf A}_i } \over {d\tau }} = 
{{e_i } \over c}\left[ {{\bf v}_i  \times {\bf B}_i } \right] - {{e_i } \over c}{{\partial
{\bf A}_i } \over {\partial \tau }} - {b \over c}\nabla _i V_i   \cr 
  & {\rm       } + \sum\limits_{k \ne i}^n {\left\{ {{{e_k } \over c}\left[ {({\bf v}_k 
\cdot \nabla _i ){\bf A}_k  +  {\bf v}_k  \times (\nabla _i  \times {\bf A}_k )} \right] -
{b \over c}\nabla _i V_k } \right\}}.  \cr}  
\eqno{(5.62)}   
$$ 

From, (5.46) and (5.47) we see that  $ ({\bf v}_k  \cdot \nabla _i ){\bf A}_k  =  - ({\bf
v}_k  \cdot \nabla _k ){\bf A}_{ik} , $ etc, so we may write (5.62) in the form (using 
$ {\bf E}_i  =  - (1/b)(\partial {\bf A}_i /\partial \tau ) - \nabla _i \Phi _i {\rm ,  }{\bf
B}_i  = {\bf v}_i  \times {\bf A}_i  $ )  
$$ 
\eqalign{
  & {c \over b}{{d\pi _i } \over {d\tau }} = e_i {\bf E}_i  + {{e_i } \over b}\left[ {{\bf
v}_i  \times {\bf B}_i } \right]  \cr 
  & {\rm       } - \sum\limits_{k \ne i}^n {\left\{ {{{e_k } \over b}\left[ {({\bf v}_k 
\cdot \nabla _k ){\bf A}_{ik}  +  {\bf v}_k  \times (\nabla _k  \times {\bf A}_{ik} )}
\right] - e_k \nabla _k \Phi _{ik} } \right\}} . \cr}  
\eqno{(5.63)}   
$$  
If we now set 
$ ({\bf v}_k  \cdot \nabla _k ){\bf A}_{ik}  = (d{\bf A}_{ik} /d\tau ) -
(\partial {\bf A}_{ik} /\partial \tau ),\quad {\bf B}_{ik}  = \nabla _k  \times {\bf
A}_{ik},\quad  {\bf E}_{ik}  =  - (1/b)(\partial {\bf A}_{ik} /\partial \tau ) - \nabla
_k \Phi _{ik},\quad {\rm and}\quad {\bf F}_{ik}  = e_k {\bf E}_{ik}  + (e_k /b){\bf
v}_k  \times {\bf B}_{ik},  $ 
we have  
$$ 
{c \over b}{{d\pi _i } \over {d\tau }} = {\bf
F}_i  - \sum\limits_{k \ne i}^n {\left\{ {{\bf F}_{ik}  + {{e_k } \over b}{{d{\bf
A}_{ik} }
\over {d\tau }}} \right\}} .  
\eqno{(5.64)}  
$$  
If we use  $ \pi _i  = \bar m_i {\bf u}_i,\,\bar m_i  = m_i \left[ {1 - {{V_i }
\mathord{\left/ {\vphantom {{V_i } {H_i }}} \right.
 \kern-\nulldelimiterspace} {H_i }}} \right], \, {\bf u}_i  = \left[ {{{c{\bf v}_i }
\mathord{\left/ {\vphantom {{c{\bf v}_i } {(b^2  - \mathop {\bf v}\nolimits_i^2
)^{1/2} }}} \right. \kern-\nulldelimiterspace} {(b^2  - \mathop {\bf v}\nolimits_i^2
)^{1/2} }}} \right],$ we get ($b$ is constant)	  
$$ 
{d \over {d\tau }}\left( {{{\tilde m_i {\bf v}_i } \over
{\sqrt {1 - \left( {{{\mathop {\bf v}\nolimits_i^2 } \mathord{\left/
 {\vphantom {{\mathop {\bf v}\nolimits_i^2 } {b^2 }}} \right.
 \kern-\nulldelimiterspace} {b^2 }}} \right)} }}} \right) = {\bf F}_i  - \sum\limits_{k
\ne i}^n {\left\{ {{\bf F}_{ik}  + {{e_k } \over b}{{d{\bf A}_{ik} } \over {d\tau }}}
\right\}} ,{\rm   }\tilde m_i  = \left( {{c \over b}} \right)^2 \bar m_i . 
\eqno{(5.65)} 
$$

In order to interpret equation (5.65), we return to equation (5.54) and use the fact that 
$ (1/b_i )(\partial /\partial \tau _i ) = (1/b)(\partial /\partial \tau ) $ and  $ \left( {{{{\bf
u}_i } \mathord{\left/ {\vphantom {{{\bf u}_i } {b_i }}} \right.
\kern-\nulldelimiterspace} {b_i }}} \right) = \left( {{{{\bf v}_i } \mathord{\left/
 {\vphantom {{{\bf v}_i } b}} \right. \kern-\nulldelimiterspace} b}} \right) $ to get 
$$
 - {1 \over {b_i }}{{\partial {\bf A}_i } \over {\partial \tau _i }} - \nabla _i \Phi _i  =  
- {1 \over b}{{\partial {\bf A}_i } \over {\partial \tau }} - \nabla _i \Phi _i ,\,\, {{e_i }
\over {b_i }}\left( { {\bf u}_i  \times {\bf B}_i } \right) = {{e_i } \over b}\left( {{\bf
v}_i  \times {\bf B}_i } \right).   
$$   
This means that our force ${\bf F}_i $ in (5.65) is identical to the right-hand side of
(5.53) (the local Lorentz force).  Equation (5.65) is our replacement  for the
Lorentz-Dirac equation.  The second term on the right-hand side is the necessary
dissipative term required to satisfy Newton's third law, and represents the action of the
i-th particle on all the other particles in the system.  It is important to note that this
equation contains no third-order derivatives, so that it will satisfy the standard
conditions for existence and uniqueness of solutions for initial value problems.  It will
not contain runaway solutions, nor advanced actions, etc.  Furthermore, the equation
does not depend on the structure of the particles in the system. 

{\sl We now see that the global view of particle interactions is a pure action-at a-
distance theory while from the local point of view particle interactions are mediated
by the fields (a field theory).}  
 
For future reference, we assume that the global system is interacting with an external
force, so that  ${\bf \dot U}$ is not zero.  If we differentiate the left-hand side of
(5.65), we get  ( using $ \left( {\mathop \beta \nolimits_i^2  = {{\mathop {\bf
v}\nolimits_i^2 } \mathord{\left/ {\vphantom {{\mathop {\bf v}\nolimits_i^2 } {b^2
}}} \right. \kern-\nulldelimiterspace} {b^2 }}} \right) $),  
$$ 
\eqalign{
  & {c \over b}{{d\pi _i } \over {d\tau }} = {{\tilde m_i {\bf \dot v}_i } \over {\left[
{1 - \mathop \beta \nolimits_i^2 } \right]^{1/2} }} + {{\tilde m_i {\bf v}_i \left[ {{\bf
v}_i  \cdot {\bf \dot v}_i  - {\bf U} \cdot {\bf \dot U}} \right]} \over {b^2 \left[ {1 -
\mathop \beta \nolimits_i^2 } \right]^{3/2} }}  \cr 
  &  - {{\tilde m_i {\bf v}_i } \over {\left[ {1 - \mathop \beta \nolimits_i^2 }
\right]^{1/2} }}{d \over {d\tau }}\left[ {\ln \left( {1 - {{V_i } \over {H_i }}} \right)}
\right],{\rm   }\tilde m_i  = m_i {{c^2 } \over {b^2 }}\left( {1 - {{V_i } \over {H_i
}}} \right). \cr}  
\eqno{(5.66)}  
$$   
Taking the dot product with ${\bf  v}_i $, we obtain the effective power transfer for
the i-th particle  
$$ 
\eqalign{
  & {{\tilde m_i } \over {2\left[ {1 - \mathop \beta \nolimits_i^2 } \right]^{3/2} }}{
{d\left\| {{\bf v}_i } \right\|^2 } \over {d\tau }} - {{\tilde m_i \left\| {{\bf v}_i }
\right\|^2 \left[ {{\bf U} \cdot {\bf \dot U}} \right]} \over {b^2 \left[ {1 - \mathop
\beta \nolimits_i^2 } \right]^{3/2} }} - {{\tilde m_i \left\| {{\bf v}_i } \right\|^2 }
\over {\left[ {1 - \mathop \beta \nolimits_i^2 } \right]^{1/2} }}{d \over {d\tau }}\left[
{\ln \left( {1 - {{V_i } \over {H_i }}} \right)} \right]  \cr 
  &  = {\bf v}_i  \cdot {\bf F}_i  - \sum\limits_{k \ne i}^n {\left\{ {{\bf v}_i  \cdot
{\bf F}_{ik}  +  {{e_k } \over b}\left( {{\bf v}_i  \cdot {{d{\bf A}_{ik} } \over
{d\tau }}} \right)} \right\}} . \cr} 
 \eqno{(5.67)}  
$$ 

\noindent If  ${\bf  U} = 0$, (5.64) and (5.67) become ($ \mathop \beta \nolimits_i^2 
= {{\mathop {\bf v}\nolimits_i^2 } \mathord{\left/
 {\vphantom {{\mathop {\bf v}\nolimits_i^2 } {c^2 }}} \right.
 \kern-\nulldelimiterspace} {c^2 }},\tilde m_i =  \bar m_i {\rm , and }\,\,\tau = t $) 
$$
{d \over {dt}}\left( {{{\bar m_i {\bf v}_i } \over {\sqrt {1 - \beta _i^2 } }}} \right) = 
{\bf F}_i  - \sum\limits_{k \ne i}^n {\left\{ {{\bf F}_{ik}  + {{e_k } \over c}{{d{\bf
A}_{ik} } \over {dt}}} \right\}} ,  
\eqno{(5.68)}  
$$  
$$ 
\eqalign{
  & {{\bar m_i } \over {2\left[ {1 - \mathop \beta \nolimits_i^2 } \right]^{3/2}
}}{{d\left\| { {\bf v}_i } \right\|^2 } \over {dt}} - {{\bar m_i \left\| {{\bf v}_i }
\right\|^2 } \over {\left[ {1 - \mathop \beta \nolimits_i^2 } \right]^{1/2} }}{d \over
{dt}}\left[ {\ln \left( {1 - {{V_i } \over {H_i }}} \right)} \right]  \cr 
  &  = {\bf v}_i  \cdot {\bf F}_i  - \sum\limits_{k \ne i}^n {\left\{ {{\bf v}_i  \cdot
{\bf F}_{ik}   + {{e_k } \over c}\left( {{\bf v}_i  \cdot {{d{\bf A}_{ik} } \over
{dt}}} \right)} \right\}} . \cr}   
\eqno{(5.69)}  
$$  
It follows that, even when the global system is at rest in the frame of the observer, our
theory is distinct.   In closing this section we note that summing equation $(5.65)$ or
$(5.68)$ on i gives zero as expected, reflecting conservation of the global
momentum.   

\bigskip 
\noindent {\bf 6.0  Discussion}  

\bigskip
\noindent {\bf 6.1  Proper-time of the Source} 

\smallskip 
\noindent In this paper, we have shown that  Maxwell's equations have a
mathematically equivalent formulation and additional symmetry group that fixes the
proper-time of the source for all observers.  The new group is closely related to the
Lorentz group and, in fact, at the local level, is a nonlinear and nonlocal
representation.    We have constructed a dual theory using the proper-time of the
source and have shown that it is covariant with respect to this group. However, the
speed of light now depends on the motion of the source and the new group replaces
time transformations between observers by transformations of the velocity of light
with respect to the source for different observers.  This implies that the speed of light
can be greater than its value in any fixed inertial frame.  In the new formulation, the
second postulate of the special relativity is only true when the source is in the rest
frame of the observer.  We have further shown that, for any closed system of particles,
there is a global inertial frame and unique (invariant) global proper-clock (for each
observer) from which to observe the system.  In this case, the corresponding group
differs from the Lorentz group by a scale transformation. This global proper-clock is
intrinsically related to the proper-clocks of the individual particles in the system and
provides a unique definition of simultaneity for all events associated with the system.  
Hence, at the global level, we can always choose a unique  observer-independent
measure of time for the study of physical systems.  One important consequence of this
result can be stated as a theorem. 

\bigskip 
\noindent {\bf Theorem 6.1} {\sl Suppose that the observable universe is
representable in the sense that the observed ratio of mass to total energy is constant and
independent of our observed portion of the universe.  Then the universe has a unique
clock that is available to all observers}.

\medskip 
\noindent  The above assumptions are equivalent to the homogeneity and
isotropy of the energy and mass density of the universe.

The use of a global variable without attaching physical meaning to it dates back to the
early work of Tetrode and Fock (for a review, see Fanchi$^{51}$).  However, starting
in  the 1970's, Horwitz and Piron$^{60}$ and later Fanchi$^{51}$ began to suggest 
the use of a special clock for global systems which they called the historical time. 
They predicted that such a variable should exist as a real physical parameter and 
Fanchi$^{52}$ suggested experiments to detect this clock.  In our approach we treat
the transformation from observer proper-time to global system proper-time as a
canonical (contact) transformation on extended phase space.  This approach allows us
to identify the canonical Hamiltonian and the associated Lie algebra (Poisson)
bracket.  Hence, we suggest that this global proper-time is the one sought by the above
researchers.   From an operational point of view, all observers can identify the time 
according to this (global) clock by recording the time on their clock, use the 
experimentally determine value for the velocity $W$, of the center of mass  of the
system, and then use equation $(1.3a)$. 

Rohrlich$^{61}$ has recently conducted a very interesting study of the classical
self-force for the dynamics of finite-sized particles with both electromagnetic and
gravitational self-interactions (using the Lorentz-Dirac equation).  He posits his model
as a replacement for the point-particle model which is beyond the validity of the
classical theory.  His approximations neglect the nonlinear terms in the derivatives of
the acceleration and leads to more reasonable equations of motion, but violates
time-reversal invariance.  This suggests that a successful classical theory which does
not require the point-particle concept may help to explain time-reversal noninvariance
at the macro-level.

As noted earlier, the proper-time theory does not depend on the size, structure, or
geometry of the charge distribution. Furthermore, the global fields of any system of
radiating particles in a closed domain will quickly leak radiation into every part of the
domain.    Since the field equations carry intrinsic information about the velocity and
acceleration of each particle at the moment of dissipation, any observer will only
receive information about the past behavior of the particles in the system.  Since the
observed radiation is an average over all the particles, this provides an explanation for
the arrow of time as a statistical effect as suggested by Einstein.  Also, since we only
use the retarded solutions of Maxwell's equations, we may follow the suggestion of
Feynman$^{32}$ and St\"uckelberg$^{62}$  and treat antimatter as matter with its
proper-time reversed. 

The above approach also provides us with a simple answer for questions about
conservation laws during the big bang.  If we assume that the big bang created two
separate universes, one with matter (moving forward in proper-time), and one with
antimatter (moving backward in proper-time).  Then  all global (physical) quantities in
our universe will be conserved while providing us with a nice explanation for the lack
of large concentrations of antimatter in our universe.

\bigskip
\noindent {\bf 6.2  Equivalent Theories and Convention}  
\smallskip 
\noindent It is no doubt more unsettling to many that the two theories could be
mathematically equivalent but not physically equivalent.  It is more natural to expect
that two mathematically  equivalent theories would also be physically equivalent, and
there are a number of historical examples to support such expectations; the
Lagrange-Hamiltonian formulation of classical mechanics, the
Heisenberg-Schr\"odinger formulation of quantum mechanics, and the
Feynman-Schwinger-Tomonaga formulation of quantum electrodynamics.  In the first
case, both formulations have proved equally valuable depending on the purpose. 
However, the latter two cases raise interesting questions.

After Feynman  constructed a path integral formulation of quantum mechanics, it was
shown to physically include the Heisenberg-Schrodinger formulation.  However, it
has never been shown to be mathematically  equivalent since there are well-known
serious foundational problems  with the mathematical notion of a path integral for
quantum theory.  On the other hand, it has not been shown that the
Heisenberg-Schr\"odinger formulation is physically equivalent to the Feynman path
integral approach. (There are theories where the path integral  approach is easy, while
the other two approaches are difficult to construct.)

In order to prove that the Feynman formulation of QED was physically equivalent to
the Schwinger-Tomonaga formulation, Dyson$^{63}$  assumed that time had the
additional property of an index which kept track of the time an operator operates
(time-ordering).  This represents a new physical input to theory formulation
 and has only recently received any mathematical attention$^{64,65,66}$.   Thus, 
mathematical  equivalence has not been shown, and although some progress has been
made, we are far from a solution.

In our opinion, the Feynman path integral approach is physically  more general than
that of  Heisenberg and Schr\"odinger, and his formulation of QED is physically more
general than that of Schwinger and Tomonaga.   In both cases, he introduces new 
concepts that make it physically easier to think about and solve problems.   What
Feynman did was to show that it is still possible to formulate theories which more
closely represent the way the world appears to us in our consciousness. 

It was Poincar\'e$^{67}$ who first noticed that some hypotheses (assumptions), which
are made for theory construction, arise because of  empirical data, while others occur
because they are convenient.  The convenient  hypotheses are generally imposed by
the mathematical structures we use to represent physical theories.  These hypotheses
are called  conventions by  Poincar\'e in order to point out the fact that different
conventions could lead to different theories which would be mathematically
equivalent.  He was not sure that the theories would be physically different, but he
seems to have left open that possibility. The work of this paper shows that different
conventions can lead to different physical theories.  Since all inerital reference frames
are equivalent, the one chosen by any observer is a convention.  If we seek simplicity,
we can all attach our frames to the MBR and use the proper-time of the universe for
our global clock.  In this case, we could satisfy the two postulates of the special theory,
while the field and particle equations of any system would be {\it invariant} under  the
action of the Lorentz group (for all observers).

\bigskip 
\noindent {\bf 6.3  Velocity of Light}

\smallskip 
\noindent
The price paid for the results of this paper will certainly seem high to many.  We have
rejected the third postulate of Minkowski that time be put on an equal footing  with
position and made a coordinate for four-geometry.  We have also rejected  the
assumption (convention) that  the observer proper-time be used to define the dynamics
of an observed system.  Thus, in our approach, the time is a (intrinsic) dynamical
variable which must be determined by experiment along with other properties (of the
observed system).  This leads to a new interpretive framework in which the second
postulate is only true when the source is at rest in the frame of the observer.   Thus, we
have reduced the observer reference frame to the prerelativistic three-geometry of
Euclidean space.  The observer's clock is now a part of the measuring equipment
which is used to determined the proper-time of the source.

The proper-time formulation has an obvious disadvantage since, it is generally
believed that,  all the available experimental evidence supports the second postulate of
special relativity (that the velocity of light is constant).  Einstein$^{68}$ pointed out
in a footnote to his second paper: ``The principle of the constancy of the velocity of
light is of course contained in Maxwell's equations."  What he meant by this was that
the second postulate follows from the fact that the constant $c$ in Maxwell's
equations is an invariant for all (inertial) observers.   Since that time, many
experiments have been done to verify that assumption.  However, in 1965,
Fox$^{69}$ wrote a very important paper which reviewed the evidence for constant
$c$  and against the emission theory of Ritz$^{44}$.   His conclusion was that all
previous experiments were flawed for a number of reasons. In many cases, analysis of
the experimental data failed to take into account the (now well-known) extinction
theorem of Ewald and Oseen (see Jackson$^{2}$  ).  The only data found that firmly
supported the second postulate came from experiments on the lifetime  of fast mesons
and the velocity of ${\gamma}$  rays and light from moving sources.  In his
conclusion Fox states that `` ${\cdots}$ Unless something has been overlooked, these
seem to be the only pieces of experimental evidence we have.  This is surprising in
light of the long history and importance of the problem."  These ``pieces of
experimental evidence" have another interpretation in the proper-time theory.  As
noted in Section 1, the lifetime of fast mesons is the fixed value measured when they
are at rest while their velocity is now computed using the proper-time of the meson
which is derived from the experiment.  The same  interpretation applies to
${\gamma}$  rays and light from moving sources.  Thus, the  same experiments that
support $c$  as constant when we assume that the observer proper-time should be used
to formulate the theory also supports the result that the speed of light depends on the
motion of the source when we assume that the source proper-time should be used to
formulate the theory.

\bigskip
\noindent {\bf 6.4  Photon Mass}  
\smallskip  
\noindent Work on the question of photon mass has focused on the addition of a mass
term to  the Lagrangian density for Maxwell's equations and generally leads to the
Proca equation  ( see Bargmann and Wigner$^{70}$).  Early work in this direction
can be traced back from the  paper of  Schr{\"o}dinger and Bass$^{71}$.  As in our
approach, the speed of light is no longer constant in all reference frames.  In this case,
the fields are distorted by the mass term and  experiments of Goldhaber and
Nieto$^{72}$ use geomagnetic data to set an upper bound of  $3\times
10^{-24}\,GeV$ for the mass term (see Jackiw$^{73}$ ).  This approach causes
gauge problems, and  has not found favor at the classical level.   The proper-time
theory is fully gauge invariant and the (photon) mass is dynamical, appearing only
during acceleration of the source.    

It should be recalled that Maxwell's equations are (spin $1$) relativistic wave
equations (see Akhiezer and Berestetskii$^{74}$). On the other hand, the experiments
of Pound and Snider$^{75}$ show directly that photons have an apparent weight (as
one would expect of any material object).  These experiments do not depend on either
the special or general theory of relativity and are not directly dependent on frequency
or wavelength measurements.    The existence of a small mass for the photon has
important implications for QED.  It is well-known that a small photon mass can
eliminate the infrared catastrophe (see Feynman$^{76}$).  

\bigskip \vfill\eject

\noindent{\bf Acknowledments}  
\smallskip
Work for this paper was begun while the first author was supported as a member of the
School of Mathematics in the Institute for Advanced Study, Princeton, N. J., and
completed during a visting appointment in the physics department at the  University of
Michigan. The authors would  like to acknowledge important discussions, comments
and encouragement from Professors G. Wienreich and H. Winful of the University of
Michigan, and  Professor Horwitz from the University of Tel Aviv, Isreal.  We would
like to give special thanks to Professor M. Wegener of the University of Aarhus,
Denmark for an introduction to the work of Poincar\'e.

\bigskip
\noindent {\bf Appendix}

\smallskip
\noindent
In this appendix, we outline the derivation of (3.54) from the angular distribution
(3.52) by taking the limit as $r\to \infty$ after integrating over a sphere of radius $r$.  
The integrations over the azimuthal angle $\phi $ are easily done. Then, for the
integrations over the polar angle $\theta $, it is convenient to make the change of
variable $\mu =\cos \theta $ and for $a=2,3,\ldots ,\ b=0,1,2\ldots ,$ define the
following sequence of integrals: $$ I_{a,b}\equiv \int\limits_{-1}^1 {\left( {1-\beta
\mu } \right)^{-a}}\mu ^bd\mu. \eqno(A1) $$  We then obtain from $(3.52)$ that

$$ \eqalign{&\mathop {\lim }\limits_{r\to \infty }\int\!\!\!\int {-{{dU} \over
{dt}}(\Omega )d\Omega}={{bq^2\left| {\bar {\bf a}} \right|^2} \over {\bar
b^4}}\left\{ {\left( {1-{1 \over 2}\sin ^2\alpha }\right)I_{4,0}} \right.\cr  &+\left( {{1
\over 2}\sin ^2\alpha -\cos ^2\alpha } \right)I_{4,2}-2\beta \left[ {\beta \cos ^2\alpha
\left({I_{5,0}-I_{5,2}} \right)} \right.\cr
  &\left. {+\left( {-\cos ^2\alpha +{1 \over 2}\sin ^2\alpha} \right)\left(
{I_{5,1}-I_{5,3}} \right)} \right]\cr  &+\beta ^2\left[ {\left( {\beta ^2\cos ^2\alpha
+{1 \over 2}\sin ^2\alpha } \right)\left( {I_{6,0}-I_{6,2}} \right)}\right.\cr  &+2\beta
\cos ^2\alpha \left( {I_{6,3}-I_{6,1}} \right)\ \left. {+\left. {\left( {\cos ^2\alpha -{1
\over 2}\sin ^2\alpha } \right)\left({I_{6,2}-I_{6,4}} \right)} \right]}
\right\}.\cr}\eqno(A2) $$ Relations among the integrals $(A1)$ for different integer
values of $a$ and $b$ are easily obtained by integration by parts:

$$ I_{a,b}={1 \over {\beta \left( {a-1} \right)}}\left[{\left( {1-\beta }
\right)^{-(a-1)}-\left( {-1} \right)^b\left( {1+\beta } \right)} \right]-{b \over{\beta
\left( {a-1} \right)}}I_{a-1,b-1},\eqno(A3) $$      for $a\ge 2,\ \ b\ge 1;$ and for
$b=0$, only the first term contributes:

$$ I_{a,0}={1 \over {\beta \left( {a-1} \right)}}\left[ {\left( {1-\beta }
\right)^{-(a-1)}-\left( {-1} \right)^b\left( {1+\beta } \right)} \right],\,\,a\ge
2.\eqno(A4) $$

\par We note that the differences of the integrals $(A1)$ that occur in $(A2)$ are of
the type $I_{a,b}-I_{a,b+2}$ for given values of $a$ and $b$.  For differences of this
type, the term in brackets in $(A3)$ does not contribute and we have:

$$ I_{a,b}-I_{a,b+2}={1 \over {\beta \left( {a-1} \right)}}\left[ {-b\left(
{I_{a-1,b-1}-I_{a-1,b+1}}\right)+2I_{a-1,b+1}} \right],\eqno(A5) $$  for integer
values of $a$ and $b$ such that $a\ge 3,\ \ b\ge 1$.  For $b=0$ the difference term on
the right hand side of $(A5)$ is missing  and we have:

$$ I_{a,0}-I_{a,2}={2 \over {\beta \left( {a-1} \right)}}I_{a-1,1},\,\;\,a\ge
3.\eqno(A6) $$

\par To use the above results to evaluate $(A2)$, we start with differences of the form
$(A5)$ and $(A6)$ with $a=6$ and $b=1,2$.  The terms which arise from the
difference term on the right-hand side of $(A5)$ combine with the terms with $a=5$
which are already present in $(A2)$.  After combining the coefficients of similar
terms, we can then apply the process again to the integrals $(A1)$ with $a=5$.  Now
we have a difference from the situation with the integrals involving $a=6$ that , in
addition to having differences of the form $(A5)$ with $a=5$ and $b=1$ and of
$(A6)$ with $a=5$, we also have the integrals $I_{5,1}$ and $I_{5,0}$ which are not
differences. However, these are easily evaluated by use of $(A3)$ (giving a term
involving $I_{4,0}$) and $(A4)$, respectively.  We can continue this procedure to
successively lower values of $a$, terminating at the value $a=2$.  The substitution of
the various values of $I_{a,b}$ and elimination of $\cos ^2\theta $ by use of the
identity $\cos ^2\theta =1-\sin ^2\theta $ leads to the result $(3.54)$.

\bigskip

\noindent

\vfill\eject

\frenchspacing

\noindent {\bf References} \medskip 

\smallskip\noindent \item{[1]} R. P. Feynman, R. B. Leighton, and M. Sands,  {\it
The Feynman Lectures on Physics}, Vol II.  Addison-Wesley, New York (1974).

\smallskip\noindent \item{[2]} J. D. Jackson, {\it Classical Electrodynamics}, John 
Wiley {\&} Sons, New York (1975).

\smallskip\noindent \item{[3]} H. A. Lorentz,  {\it Archives Neerlandaises des
Sciences Exactes et Naturelles},  {\bf 25}, 353 (1892).

\smallskip\noindent \item{[4]} H. A. Lorentz,  {\it The Theory of Electrons} B. G. 
Teubner, Leipzig, 1906; (reprinted by Dover, New York, 1952).

\smallskip\noindent \item{[5]} A. Einstein,  Ann. d. Phys. {\bf 17}, 891 (1905).

\smallskip\noindent \item{[6]} D. E. Spencer and U. Y.  Shama, Physics Essays {\bf
9},  476 (1996).

\smallskip\noindent \item{[7]} A. Einstein, Jahrbuch Radioaktivitat {\bf V},
 422 (1907) (Berichtigungen).

\smallskip\noindent \item{[8]} H. Poincar${\acute e}$,  C.R. Acad. Sci. Paris {\bf
140}, 1504 (1905).

\smallskip \noindent \item{[9]} H. Minkowski, Physikalische Zeitschrift {\bf 10},104
(1909).

\smallskip\noindent \item{[10]} E. Whittaker, {\it A History of Aether and 
Electricity}, Vol. I, Thomas Nelson and Sons, London,  (1951).

\smallskip\noindent \item{[11]} M. Dresden, in {\it Renormalization: From Lorentz 
to Landau (and Beyond)}, L. M. Brown (ed.), Springer-Verlag, New York,  (1993).

\smallskip\noindent \item{[12]} M.H.L. Pryce, Proc. Roy. Soc. London  A {\bf 195},
400 (1948).

\smallskip\noindent \item{[13]} P. A. M. Dirac, Rev. Mod. Phys.  {\bf 21}, 392
(1949).

\smallskip\noindent \item{[14]} H. Leutwyler and J. Stern, Ann. Phys. (N. Y.) {\bf
112}, 94 (1978).

\smallskip\noindent \item{[15]} B. Bakamjian and L. H. Thomas, Phys. Rev. {\bf 92},
1300 (1953).

\smallskip\noindent \item{[16]} D. G. Currie, T. F. Jordan, and E. C. G. Sudarshan,
Rev. Mod. Phys. {\bf 35}, 350  (1963).

\smallskip\noindent \item{[17]} E. C. G. Sudarshan and N. Mukunda, {\it Classical
Dynamics: A Modern Perspective}. John  Wiley {\&} Sons, New York (1974).

\smallskip\noindent \item{[18]} R. Fong and J. Sucher, J. Math. Phys. {\bf 5}, 456
(1964).

\smallskip\noindent \item{[19]} A. Peres, Symposia Mathematica {\bf 12}, 61 (1973).

\smallskip\noindent \item{[20]} E. P. Wigner, in {\it Aspects of Quantum Theory}, in
Honor of P. A. M.  Dirac's 70th Birthday, edited by A. Salam and E. P. Wigner
(Cambridge Univ. Press, London, 1972).

\smallskip\noindent \item{[21]} F. Rohrlich,{\it Classical Charged Particles:
Foundations of Their Theory }, Addison-Wesley, Reading, Massachusetts (1965).

\smallskip\noindent \item{[22]} S. Parrott, {\it Relativistic Electrodynamics and
Differential Geometry}, Springer, New York (1987).

\smallskip\noindent \item{[23]} F. Rohrlich, Phys. Rev. D {\bf 58}, 116002 (1999).

\smallskip\noindent \item{[24]} W. K. H. Panofsky and M. Phillips, {\it Classical
Electricity and Magnetism}, (Second Edition), Addision-Wesley, Reading, MA.
(1962).

\smallskip\noindent \item{[25]} J. A. Wheeler and R. P. Feynman,  Rev. of Mod.
Phys. {\bf 21}, 425 (1949).

\smallskip\smallskip\noindent \item{[26]} J. Schwinger, Found. Phys. {\bf 13},  2573
(1998).

\smallskip\noindent \item{[27]} M. Born and L. Infield, Proc. R. Soc. London {\bf
A144}, 425 (1934).

\smallskip\noindent \item{[28]} P. A. M. Dirac, Proc. R. Soc. London {\bf A167},
148 (1938).

\smallskip\noindent \item{[29]} F. Bopp, Ann. d. Phys. {\bf 42}, 573 (1942).

\smallskip\noindent \item{[30]} N. Rosen, Phys. Rev.  {\bf 72}, 298 (1947).

\smallskip\noindent \item{[31]} B. Podolsky and P. Schwed, Rev. Mod. Phys.  {\bf
20}, 40 (1948).

\smallskip\noindent \item{[32]} R. P. Feynman, Phys. Rev.  {\bf 74}, 939 (1948).

\smallskip\noindent \item{[33]} R. Haag, Z. f. Naturf.  {\bf 10a}, 752 (1955).

\smallskip\noindent \item{[34]} S. Parrott and D. J. Endres,  Found. Phys. {\bf 25},
441 (1995).

\smallskip\noindent \item{[35]} F. E. Low,  Ann. Phys. {\bf 266}, 274 (1998).

\smallskip\noindent \item{[36]} A. A. Penzias and R. W. Wilson,  Ap. J. {\bf 142}, 
419 (1965).

\smallskip\noindent \item{[37]}  P.J.E. Peebles, {\it Principles of Physical 
Cosmology},  Princeton University Press, London (1993).

\smallskip\noindent \item{[38]} S. Schweber, {\it QED And The Men Who Made It
 },  Princeton University Press, London (1994).

\smallskip\noindent \item{[39]} B.  French and V. Weisskopf, Phys. Rev.  {\bf 75}, 
1240 (1949).

\smallskip\noindent \item{[40]} N.  Kroll and W. Lamb, Phys. Rev.  {\bf 75},  388
(1949).

\smallskip\noindent \item{[41]} P. A. M.  Dirac, Sci. Amer.  {\bf 208}, 45 (1963).

\smallskip\noindent \item{[42]} T. P. Gill, {\it The Doppler Effect} , Logos Press,
London (1965).

\smallskip\noindent \item{[43]} G. A. Schott, Phil.  Mag. {\bf 29}, 49 (1915).

\smallskip\noindent \item{[44]} W. Ritz, Archives des Sciences Physiques et
Naturelles {\bf 16}, 209 (1908).

\smallskip\noindent \item{[45]} C. Moller, {\it The Theory of Relativity}, Oxford
Clarendon Press, London  (1960).

\smallskip\noindent \item{[46]} C. H. Papas, {\it Theory of Electromagnetic Wave
Propagation}, Dover Press, New York (1988).

\smallskip\noindent \item{[47]} R. Courant and D. Hilbert, {\it Methods of
Mathematical Physics}, vol. II , Wiley-Interscience, New York (1965).

\smallskip\noindent \item{[48]} T. L. Gill and J. Lindesay, Inter. J. Theor. Phys. {\bf
32}, 2087 (1993).

\smallskip\noindent \item{[49]} T. L. Gill, Fermilab-Pub-82/60-THY.

\smallskip\noindent \item{[50]} J. P. Aparicio, F. H. Gaioli, and E. T. Garcia-Alvarez,
Phys. Rev. A {\bf 51}, 96 (1995).

\smallskip\noindent \item{[51]} J. R. Fanchi, {\it Parametrized Relativistic  Quantum
Theory}, Kluwer, Dordrecht, (1993).

\smallskip\noindent \item{[52]} J.R. Fanchi,  Found. Phys. {\bf 23}, 487 (1993).

\smallskip\noindent \item{[53]} E. P. Wigner, Ann. Math. {\bf 40}, 149 (1939).

\smallskip\noindent \item{[54]} G. L. Strobel, Inter. J. Theor. Phys. {\bf 37}, 2087
 (1998).

\smallskip\noindent \item{[55]} T. L. Gill, W.W. Zachary, and J. Lindesay, Inter. J. 
Theor. Physics {\bf 37}, 2573 (1998).

\smallskip\noindent \item{[56]} P. A. M.  Dirac, V. A. Fock, and B. Podolsky, Phys.
Z. Sowj. Un. {\bf 2}, 6 (1932) [Reprinted in J. Schwinger, ed. {\it Selected Papers in
Quantum Electrodynamics, Dover, New York (1958)}].

\smallskip\noindent \item{[57]} F. Rohrlich and L. P. Horwitz, Phys. Rev. D {\bf 24},
1528 (1981).

\smallskip\noindent \item{[58]} G. Longhi, L. Lusanna, and J. M. Pons, J. Math. 
Phys. {\bf 30}, 1893 (1989).

\smallskip\noindent \item{[59]} R. J. Hughes, Amer. J. Phys. {\bf 60}, 301 (1992).

\smallskip\noindent \item{[60]} L. P. Horwitz and C. Piron, Helv. Phys. Acta {\bf
46}, 316 (1981).

\smallskip\noindent \item{[61]} F. Rohrlich, Phys. Rev. D {\bf 60}, 084017 (1999).

\smallskip\noindent \item{[62]} E. C. G. St\"uckelberg, Helv. Phys. Acta {\bf 15}, 23
(1942).

\smallskip\noindent \item{[63]} F. J. Dyson, Phys. Rev. D {\bf 75}, 486, 1736 (1949).

\smallskip\noindent \item{[64]} G. W. Johnson and M. L. Lapidus, Mem. Amer.
Math.  Soc. {\bf 62}, 1 (1986).

\smallskip\noindent \item{[65]} T. L. Gill and W.W. Zachary, J. Math. Phys. {\bf 28},
1459
 (1987).
 
\smallskip\noindent \item{[66]} G. W. Johnson and M. L. Lapidus, {\sl The  Feynman
Integral and Feynman's Operational Calculus}, Oxford U. Press,  New York, (2000).

\smallskip\noindent \item{[67]} H. Poincar\'e, {\sl Science and Hypothesis},  Dover
Press, New York, (1952).

\smallskip\noindent \item{[68]} A. Einstein,  Ann. d. Phys. {\bf 18}, 639 (1905).

\smallskip\noindent \item{[69]} J. G. Fox, Amer. J. Phys. {\bf 33}, 1 (1965).

\smallskip\noindent \item{[70]} V. Bargmann and E. P. Wigner, Proc. Nat. Acad. Sci.
{\bf 34}, 211 (1948).

\smallskip\noindent \item{[71]} E. Schr\"odinger and L. Bass, Proc. R. Soc. London
{\bf A232}, 1 (1938).

\smallskip\noindent \item{[72]} A. Goldhaber and M. Nieto,  Rev. Mod. Phys. {\bf
43}, 277 (1971).

\smallskip\noindent \item{[73]} R. Jackiw, Comments Mod. Phys. {\bf 1A}, 1 (1999).

\smallskip\noindent \item{[74]} A. I. Akhiezer and V. D. Berestetskii, {\it Quantum
Electrodynamics}, Wiley-Interscience, New York (1965) 

\smallskip\noindent \item{[75]} R. V. Pound and J. L. Snider, Phys. Rev.  {\bf 140},
B788 (1965).

\smallskip\noindent \item{[76]} R. P. Feynman, {\it Quantum Electrodynamics}, W.
A. Benjamin, New York  (1964)

\vfill

\bye

\end